\documentclass[preprint,12pt]{elsarticle}
\usepackage{graphics} % for pdf, bitmapped graphics files
\usepackage{graphicx,caption}
\usepackage{mathtools}
\usepackage{tikz}
\usepackage{mathrsfs}
\usepackage{bm}
\usepackage{amsbsy}
\usepackage{amsfonts}
\usepackage{amssymb}
\usepackage{amsmath}% mathtools includes this so this is optional
\usepackage{soul,color}
\usepackage{booktabs}
\usepackage{subfiles}
\usepackage{amsmath}
\usepackage{lscape}

\journal{Engineering Applications of Artificial Intelligence}

\begin{document}
	
\begin{frontmatter}

\title{A Survey on Reinforcement Learning in Aviation Applications}

\author[a]{Pouria Razzaghi}
\author[a]{Amin Tabrizian}
\author[a]{Wei Guo}
\author[a]{Shulu Chen}
\author[a]{Abenezer Taye}
\author[a]{Ellis Thompson}
\author[b]{Alexis Bregeon}
\author[c]{Ali Baheri}
\author[a]{Peng Wei}

\affiliation[a]{organization={George Washington University},%Department and Organization
            city={Washington D.C.},
            country={USA}}

\affiliation[b]{organization={Ecole Nationale de l'Aviation Civile},
            city={Toulouse},
            country={France}}

\affiliation[c]{organization={Rochester Institute of Technology},%Department and Organization
            city={Rochester, NY},
            country={USA}}

\begin{abstract}
    Reinforcement learning (RL) has emerged as a powerful tool for addressing complex decision making problems in various domains, including aviation. This paper provides a comprehensive overview of RL and its applications in the aviation industry. We begin by introducing the fundamental concepts and algorithms of RL, highlighting their unique advantages in learning from interaction and optimizing decision-making processes. We then delve into a detailed examination of the successful implementation of RL methods in aviation, covering areas such as flight control, air traffic management, airline revenue management, aircraft maintenance scheduling, etc. Furthermore, we discuss the potential benefits of RL in enhancing safety, and sustainability within the aviation sector. Finally, we identify and explore open challenges and areas for future research, emphasizing the need for continued innovation and collaboration between the fields of reinforcement learning and aviation.
\end{abstract}

\end{frontmatter}

\section{Introduction}
Reinforcement learning (RL) has shown promising performances in complex sequential decision making problems. Compared with classical decision making methods such as control and optimization, RL takes advantages of the availability of large-scale datasets, fast-time simulators, state-of-the-art neural network architectures, and high-performance computing resources to build models and algorithms. In addition, RL methods demonstrate scalability, run-time efficiency, and generalizability in stochastic, non-linear and dynamic decision problems. Many of these problems can be found in aviation and aeronautical applications, ranging from the planning problems such as airline maintenance, air traffic flow management, crew scheduling and aircraft routing, to control problems such as aircraft sequencing and separation, collision avoidance, and flight adaptive control. 

\paragraph{Significance and contributions} This paper provides a timely comprehensive survey of reinforcement learning in aviation applications. The objective of this work is to identify state-of-the-art RL algorithms in most successful aviation applications. In addition, we also identify the technical gaps in applying RL in the aviation sector, including the certification concerns of implementing learning-based, neural-network-in-the-loop RL models in safety-critical applications such as aircraft collision avoidance. We expect this paper will serve as an introductory reading material for researchers who plan to start working in this interdiscipline area of RL for aviation. 

% \hl{Peng, please write the introduction section here. This is the reviewer's comment for this section: I suggest that the authors include an Introduction section focusing on introducing this work, such as its significance, an overview of aviation, the scope of this work, and its main contributions or findings.}
\section{Preliminaries}

\subsection{Reinforcement Learning in Aviation}
Compared with model-based control and optimization methods, RL provides a data-driven, learning-based framework to formulate and solve sequential decision-making problems. Reinforcement learning frameworks have become promising due to largely improved data availability and computing power in the aviation industry. Many aviation applications can be formulated or treated as sequential decision-making problems. Some of them are offline planning problems, while others need to be solved in an online manner and are safety-critical. In this survey paper, we first describe standard RL formulations and solutions. Then we survey the landscape of existing RL-based applications in aviation. Finally, we summarize the paper, identify the technical gaps, and suggest future directions of RL research in aviation.

\begin{figure}[ht!]
	\centering
	\includegraphics[width= 1 \textwidth]{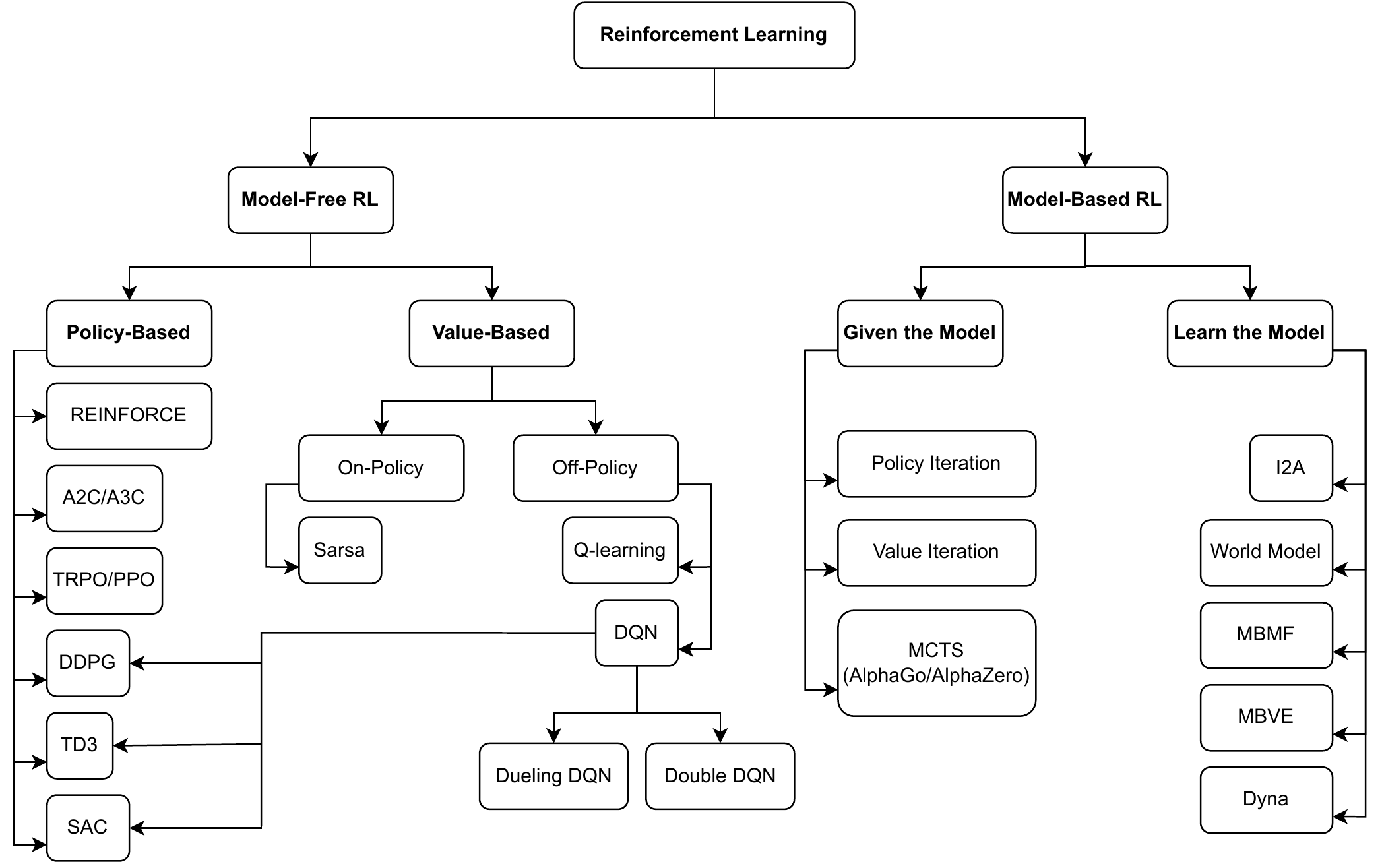}
	\caption{Categories of the reinforcement learning models and algorithms.}
	\label{RLCat}
\end{figure}

The RL models and algorithms are comprehensively outlined in the remainder of this section. To begin, we briefly describe the RL problem formulation and a few key concepts. Following that, two classical categories of RL algorithms will be presented: value-based and policy-based leanings. We then will present the more advanced techniques as well as modern actor-critic methods and multi-agent reinforcement learning (MARL). The summarized categories of the RL models and algorithms are shown in Fig.~\ref{RLCat}.

\subsection{Overview of Reinforcement Learning}
\begin{figure}[!ht]
	\centering
	\includegraphics[width= 0.6 \textwidth]{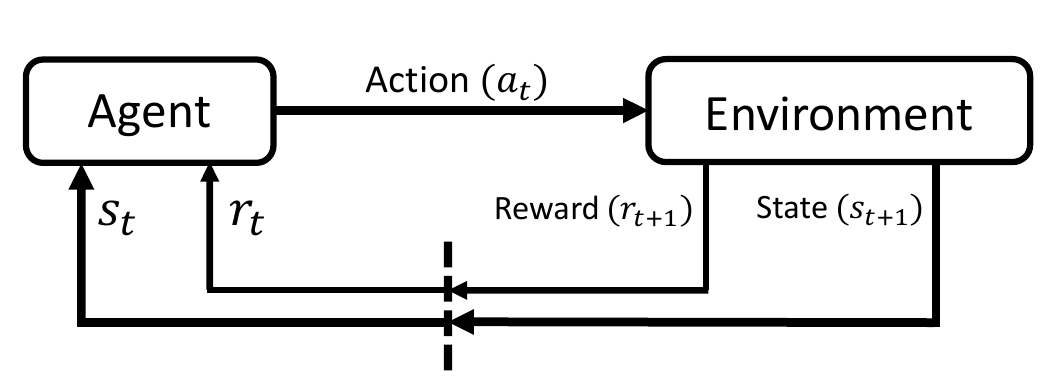}
	\caption{In RL, an agent selects an action $a_t$ based on its current state $s_t$, then it will receive a reward from the environment $r_t$ and arrive to the next state $s_{t+1}$. This process will continue until the agent arrives at a terminal state if any.}
	\label{Fig1}
\end{figure}

Reinforcement learning is a branch of machine learning that is about an agent interacting with an environment to complete a task or achieve a goal. The environment is stated in the form of a Markov decision process (MDP) used to solve sequential decision-making problems. In an MDP problem, an agent takes a series of actions to maximize the total received reward from an unknown environment. This problem can be represented by a tuple of $(\mathcal{S}, \mathcal{A}, P, R,\gamma)$, where $\mathcal{S}$ is the set of states, $\mathcal{A}$ is the set of actions, \textit{P} is the transition probability function $\left(P\left(s_{t+1}|s_t,a_t\right)\right)$ that maps a state-action ($s_t,a_t$) pair to distribution of next possible states, \textit{R} is the received reward at each step, and $\gamma$ is the discount factor representing the relative importance of future and immediate rewards. The policy, $ \pi(.) $, represents a mapping from an agent's state to a distribution on the action space. The optimal policy, $ \pi^{\ast}(.)$, takes place where the summation of expected rewards, \big($\sum_{i=0}^{\infty} \gamma^i r^{t+i}$\big), for the course of action is maximized. Fig.~\ref{Fig1} depicts a block diagram of an RL process. An agent observes its current state and reward from the environment; then, the agent selects an action according to its policy. This will change the state of the environment and the new reward and state in the next time step will be pushed back to the agent.

% If the model, $P\left(s_t+1|s_t, a_t\right)$, and reward $r_t$ are known, dynamic programming (DP) is one method for determining the optimal policy \cite{bertsekas2012dynamic} which is part of model-based RL algorithms. In the absence of a model, the agent should learn the optimal policy by observing past interactions or by directly interacting with the environment, which is model-free RL (see details in Fig.~\ref{RLCat}). In recent decades, variations and improvements have been made to methods designed to solve real-world problems using RL formulations and solution algorithms. 

One of the most important differentiation points in the RL algorithms is whether the agent has access to the model or not. Model-based RL algorithms either have direct access or can use a learned model of the environment, i.e., the agent knows $P\left(s_{t+1}|s_t, a_t\right)$, and reward $r_t$, while those algorithms that do not consider the environment model are known as model-free. In model-free RL, the agent should learn the optimal policy by observing past interactions or by directly interacting with the environment (see details in Fig.~\ref{RLCat}).
% Although model-free methods do not benefit from the potential gains in sample efficiency that may come from using a model, they are also more straightforward to implement and tune. 

There are a few model-based RL algorithms such as policy iteration, value iteration \cite{bertsekas2012dynamic}, world models \cite{ha2018world}, and model-based value expansion (MBVE) \cite{feinberg2018model}. Some algorithms have a combination of both model-free and model-based RL like imagination-augmented agents (I2A) \cite{racaniere2017imagination}, and model-based RL with model-free fine-tuning (MBMF) \cite{nagabandi2018neural}. On the contrary, model-free methods have been extensively developed and used recently. In the following, we describe some traditional and modern model-free RL methods.

\subsection{Standard Formulations of Model-Free Reinforcement Learning}
\subsubsection{Value-based Methods}
 Q-learning is one of the fundamental value-based RL algorithms introduced by Watkins \cite{watkins1989learning} at the end of the 1980s. A Q-value for every combination of state and action pair in an environment can be defined as Eq.~\ref{qval}. It represents an expected value of the cumulative reward at time step $t$ for an action ($a$) when it follows a policy $\pi$ as follows:
\begin{equation} 
	Q_{\pi}(s,a)= \mathbb{E}_{\pi}\left[ \sum_{i=0}^{\infty} \gamma^i r_{t+i+1}| s,a  \right]
	\label{qval}
\end{equation}
where $i$ is the number of steps forward at time step $t$. After updating the Q-value, the algorithm attempts to determine how valuable it is to take a particular action in a specific state. A Q-table is made by all the stored Q-values of each state-action pair in a discrete space (a Q-function approximator is used for a continuous state space-action). The policy $\pi(s) = \text{argmax}\,Q(s,a)$ yields the highest total reward. An agent selects an action to explore the environment (so-called exploration visiting almost all the state-action pairs a sufficient number of times) and observes the outcome. The Q-value can be updated by the temporal difference (TD) technique \cite{sutton2018reinforcement}:
\begin{equation}
	Q(s_t,a_t) \gets Q(s_t,a_t) + \alpha \Big(r_t+\gamma \underset{a_{t+1} \in \mathcal{A}}{\text{max}} Q(s_{t+1},a_{t+1})-Q(s_t,a_t)\Big)
	\label{qup}
\end{equation} 
where $Q' = r_t+\gamma \underset{a_{t+1} \in \mathcal{A}}{\text{max}} Q(s_{t+1},a_{t+1})$ is considered as the temporal difference target and $\alpha$ denotes the learning rate. We should note that Eq.~\ref{qval} is a stochastic approximation scheme for the Bellman optimality equation solution and it will converge to $Q^{*}$ under certain assumptions \cite{tsitsiklis1994asynchronous, qu2020finite}. 

An off-policy method learns the value of the optimal policy independent of the agent's actions. Q-learning is considered an off-policy learning algorithm since it involves updating Q-value based on experiences that are not always generated from the derived policy. Whereas state–action–reward–state–action (SARSA) is an on-policy one that generates experiences using the derived policy. For example, SARSA uses $Q' = r_t+\gamma Q(s_{t+1},a_{t+1})$ where $a_{t+1}$ is an action generated from the current policy or a given default policy.

A Monte Carlo method can be also used to estimate expected returns in non-Markovian episodic settings by averaging the results of multiple roll-outs. The Monte Carlo and TD methods have been joined and constructed the $\text{TD}(\lambda)$ \cite{sutton2018reinforcement}. The major problem of the traditional methods \cite{watkins1992q, rummery1994line} is the ``curse of dimensionality''. These methods rely on storing all the state-action pairs and representing them in a tabular format that will grow exponentially as a factor of the number of states. One approach to solving this problem is to use a deep neural network (DNN) to approximate a parameterized Q-function. This creates a deep Q-networks (DQNs) \cite{mnih2015human}. DQN introduces replay memory and a separate target network, to overcome the problem of the instability and divergence issues in the training process of the approximation. To improve the stability of learning, this method uses a separate network $\hat{Q}$ for generating targets. A specific number of iterations is fixed for each episode. Moreover, by storing all transition experiences $(s_t, a_t, s_{t+1},a_{t+1})$, the experience replay makes the random sampling for Q-learning updates more efficient \cite{lin1992self}. In addition, the DQN performance is further improved by several notable variants, such as continuous DQN (cDQN) \cite{gu2016continuous}, double DQN \cite{van2016deep}, dueling DQN \cite{wang2016dueling}, and quantile regression DQN (QR-DQN) \cite{duan2021distributional}.

\subsubsection{Policy-based Methods}
Other families of RL algorithms are policy gradient algorithms, which do not calculate value but attempt to determine an optimal policy directly. In these algorithms, a probability distribution over a set of actions ($\pi(a|s,\theta)$) concerning a policy defined as a function of parameters $\theta$ will be produced. An agent's likelihood of visiting state $s$ after applying a policy $\pi$ is described by the discounted state distribution. Using gradient ascent, the policy is optimized for the objective function:
% \begin{equation}
% 	J(\theta)= \int_{S} \rho_{\pi}(s)\, r(s, \mu_{\theta}(s)) \, ds = \mathbb{E}_{s_{\rho_{\pi}}}[r(s,\mu_{\theta}(s)) ]
% \end{equation}
\begin{equation}
	J(\theta)= \int_{S} \rho_{\pi}(s)\, r(s, \pi_{\theta}(s)) \, ds = \mathbb{E}_{s_{\rho_{\pi}}}[r(s,\pi_{\theta}(s)) ]
\end{equation}
where $\rho_{\pi}$ is the discounted state distribution \cite{sutton1999policy}. The gradients are calculated $(\theta \leftarrow \theta+\eta \triangledown J(\theta)$, where $\eta$ is the step size) while the actions are taken following the policy, and rewards are observed. More straightforwardly, the policy gradient methods choose actions directly from a model and then update the model weights to maximize the expected returns. The original policy-based method is called REINFORCE \cite{williams1992simple}, which collects a full trajectory and then updates the policy weights in the Monte Carlo style and indicates that the total return is sampled from the entire trajectory.

In the deterministic policy gradient (DPG) \cite{silver2014deterministic}, instead of using a stochastic policy $(\pi(s,\theta))$, the actions are deterministically selected using policy $\mu(s,\theta)$. DPGs are limited cases of stochastic gradient policies when the variance becomes zero.
The major drawback of a deterministic policy is the lack of exploration. For a proper exploration of the environment, the noise needs to be added and the policy becomes stochastic again \big(adding Gaussian noise $\xi, a = \pi_{\theta}(s) + \xi$\big). DPGs are therefore commonly implemented as actor-critic methods to allow off-policy exploration. Consequently, it is possible to add noise to action outputs for additional exploration without the need for a stochastic policy. Over action-value modeling, policy parametrization has the advantage of incorporating knowledge into the learning system in the form of the policy. Deep deterministic policy gradient (DDPG) \cite{lillicrap2015continuous} is a model-free off-policy algorithm for learning continuous actions, which combines ideas from DPG and DQN.

One of the problems of policy-based methods is in the gradient update. The policy performance drops if the updated policies deviate largely from previous ones. Trust region policy optimization (TRPO) \cite{schulman2015trust} ensures a monotonic improvement in policy performance by optimizing a surrogate objective function. The policy gradient updates are enforced by approximating the Kullback-Leibler (KL) divergence between the old and new policies using a quadratic approximation to be in a given range. Proximal policy optimization (PPO) \cite{schulman2017proximal} achieves the same benefits as TRPO with a simplified implementation and improved sample complexity. It is revised based on TRPO but only uses first-order optimization.

It is worth mentioning that there is no specific way to differentiate and easily define the clusters of RL methods. Most of the aforementioned methods can be pointed out as actor-critic architecture as it is illustrated in Fig.~\ref{RLCat}.   

\subsubsection{Actor-critic Methods}
The actor-critic algorithm is an eminent and widely used architecture combining policy-based and value-based methods, inheriting their advantages \cite{ke2017lightgbm}. The actor-critic algorithm can be considered a TD learning method that represents the policy function independent of the value function. It introduces the eponymous components: the actor and the critic; the policy used to select actions is called the \textit{actor}, and the estimated value function known as the \textit{critic} criticizes the actions made by the actor \cite{sutton2018reinforcement}.

The actor-critic methods achieved great success in many complex tasks; however, they suffer from various problems such as high variance, slow convergence, and local optimum. Hence, many variants have been developed to improve the performance of actor-critic methods. 
Asynchronous advantage actor-critic (A3C) \cite{mnih2016asynchronous} uses advantage estimates rather than discounted returns in the actor-critic framework and asynchronously updates both the policy and value networks on multiple parallel threads of the environment. The parallel independent environments stabilize the learning process and enable more exploration.
Advantage actor-critic (A2C) \cite{wang2016learning}, the synchronous version of A3C, uses a single agent for simplicity or waits for each agent to finish its experience to collect multiple trajectories. This modification can significantly reduce the variance of the policy gradient estimate without changing the expectation. In this method, multiple actors are trained in parallel with different exploration policies, then the global parameters get updated based on all the learning results and synchronized to each actor.
Soft actor-critic (SAC) \cite{haarnoja2018soft} with stochastic policies is an off-policy deep actor-critic algorithm based on the maximum entropy RL framework. It benefits from adding an entropy term to the reward function to encourage better exploration.
\subsection{Multi-agent Reinforcement Learning Methods}
There are some newborn control tasks to regulate the behavior of a multi-agent system interacting in a common environment. Multi-agent reinforcement learning (MARL) will be critical for the development of communication skills and other intellectual capacities, as well as for teaching agents how to cooperate without causing harm to each other. These challenging tasks motivate researchers to use multi-agent RL frameworks \cite{gronauer2022multi}. A summary of related algorithms and theories is outlined in \cite{zhang2021multi}. In the MARL framework, a set of $N$ agents interact with the same environment. At each time step and for a given state, each agent takes its action, receiving a reward. The system then propagates to the next state. 
In the MARL framework, multi-task and partial observation are usually considered \cite{omidshafiei2017deep}. The centralized and decentralized multi-agent RL methods attract much attention in aviation applications \cite {wang20213m}. One popular variant involves each agent adopting a policy, which determines the action based on local observations. As only local observations are required for the execution, this method permits decentralized implementation. However, centralized training is still required since the system's state transition relies on the actions of every agent.
Here are some popular MARL methods:
\begin{itemize}
  \item \textbf{Independent Q-Learning (IQL)}:  Each agent learns its own Q-function independently, treating other agents as part of the environment. This approach is simple but can lead to non-stationarity issues because the environment appears to change as other agents learn \cite{tan1993multi}.
 \item \textbf{Joint Action Learning (JAL)}: Agents learn a joint action-value function that depends on the actions of all agents. This approach can handle the non-stationarity problem but becomes infeasible as the number of agents and action spaces grows \cite{li2023ace}.
\item \textbf{Cooperative Multi-Agent Q-Learning}: In this approach, agents cooperate to learn a joint policy that maximizes the total reward for the group. This is often used in scenarios where agents share a common goal \cite{matignon2007hysteretic}.
 \item \textbf{Counterfactual Multi-Agent (COMA) Policy Gradients}: COMA uses a centralized critic to estimate the Q-function and decentralized actors to optimize the policies of individual agents. It addresses the credit assignment problem by using a counterfactual baseline that marginalizes a single agent's action while keeping others fixed \cite{foerster2018counterfactual}.
 \item \textbf{Multi-Agent Deep Deterministic Policy Gradient (MADDPG)}: An extension of the DDPG algorithm to multi-agent settings, where each agent has its own actor-critic pair. The critic is augmented with extra information about the policies of other agents, enabling centralized training with decentralized execution \cite{li2019robust}.
 \item \textbf{Mean Field Reinforcement Learning}: This approach is used for large-scale MARL problems, where interactions with other agents are approximated by interactions with an average agent or a mean field \cite{yang2018mean}.
 \item \textbf{Value Decomposition Networks (VDN)}: VDN decomposes the joint value function into individual value functions for each agent, which are then combined additively. This allows for efficient learning while maintaining decentralized policies \cite{sunehag2017value}.
 \item \textbf{QMIX}: An extension of VDN, QMIX uses a mixing network to combine individual value functions in a more complex, non-linear way, subject to a monotonicity constraint. This allows for better representation of joint action values while still enabling decentralized policies \cite{rashid2020weighted}.
\end{itemize}
Each of these methods has its strengths and weaknesses, and the choice of method depends on the specific requirements of the multi-agent system, such as the level of cooperation, communication constraints, and the size of the action and state spaces.
\section{Selected Applications of RL in Aviation}
\begin{figure}[!ht]
	\centering
	\includegraphics[width=0.6\columnwidth]{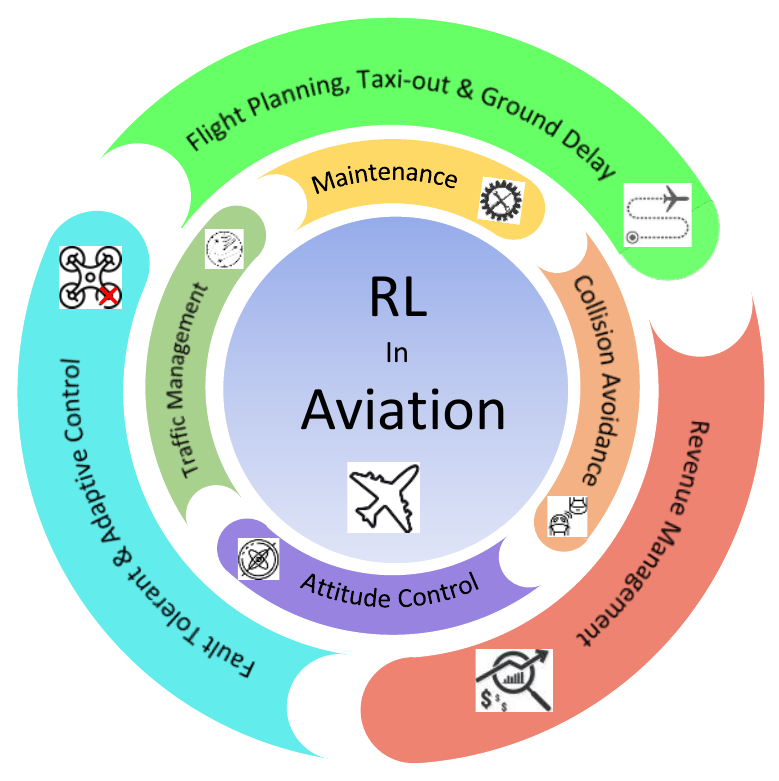}
	\caption{Taxonomy layout of RL in aviation. Different applications are shown with their corresponding illustrations. Flight planning represents a pre-defined path from the initial point. The revenue management illustration shows an increase in the profit of the airline. Controlling a drone lost one of its motors goes under the adaptive control of an air vehicle. A gimbal shape represents the attitude control of the system. The traffic management sketch depicts the control room monitor to supervise the traffic in the air. A collision avoidance picture is an alarm of avoiding a conflict between two vehicles.   }
	\label{Taxonomy}
\end{figure}

With the increasing complexity of airspace and the growth in air traffic, traditional control, optimization, and other decision making methods are being stretched to their limits. Many challenging problems in aviation can now be addressed using data-driven and machine-learning-based methods due to the availability of aviation data and significant increases in computational power. Here is a list of some of these problems: air traffic management \cite{schmidt2017review}, aircraft sequencing \cite{ahmed2018cooperative}, air traffic flow management \cite{conde2016trajectory}, taxi-out time prediction \cite{lee2016taxi}, flight delay prediction \cite{takeichi2017prediction,choi2016prediction}, trajectory prediction \cite{ayhan2016aircraft}, and aircraft performance parameter predicting \cite{alligier2015machine}. In the following, we present a comprehensive description of each application and RL's role in providing the solution for their corresponding problems.
\begin{itemize} 
    \item Collision avoidance and separation assurance: it ensures that aircraft maintain a safe distance from each other and obstacles both in the air and on the ground.  RL methods have emerged as a promising approach to enhance these safety mechanisms by enabling adaptive, real-time decision-making. Collision avoidance systems are designed to prevent aircraft from coming too close to each other or obstacles. These systems typically provide pilots or automated systems with warnings or recommended actions to maintain safe separation. RL can be used to develop more sophisticated collision avoidance algorithms that learn from experience and can adapt to a wide range of scenarios. Separation assurance involves maintaining a minimum safe distance between aircraft in all phases of flight. Air traffic controllers play a crucial role in separation assurance by directing aircraft along safe paths. RL can assist in this process by providing decision support tools that learn from past traffic patterns and optimize separation strategies in real time. This can help controllers manage traffic more efficiently, reduce the risk of close encounters, and minimize delays.
    \item Air traffic flow management: it ensures the safe and efficient movement of aircraft in the airspace and at airports. It encompasses a range of services and functions, including air traffic control, airspace management, and flow management. With the increasing complexity and volume of air traffic, maintaining efficiency and safety has led to the exploration of RL techniques. RL can be applied to manage the flow of air traffic in congested areas or during peak times, by learning to balance demand and capacity and minimize delays. In response to unforeseen events such as weather disruptions or emergencies, RL-based systems can dynamically reroute aircraft to ensure safety and minimize disruptions to the overall traffic flow. The integration of RL into air traffic management has the potential to enhance the adaptability, efficiency, and safety of the airspace system.
    \item Airline revenue management: it involves the use of sophisticated strategies to optimize the pricing and allocation of airline seats to maximize revenue. With the complexity of factors influencing ticket pricing, such as demand fluctuations, competition, and operational costs, RL models offer a promising solution by enabling dynamic and adaptive decision-making. Key challenges include accurately forecasting demand, dynamically adjusting prices based on market conditions, and balancing short-term gains with long-term profitability. RL can enhance airline revenue management by learning optimal pricing and inventory control policies through trial and error, using feedback from the market. By continuously interacting with the environment (i.e., the market), an RL agent can learn to make decisions that maximize cumulative revenue over time. Despite its potential, the application of RL in airline revenue management comes with challenges, such as the need for large amounts of data, computational complexity, and the requirement for robust and safe exploration strategies in a highly competitive market. Additionally, the integration of RL with existing revenue management systems and processes requires careful consideration.
    \item Aircraft flight and attitude control: it involves maintaining the stability and desired trajectory of an aircraft during flight and is responsible for managing the orientation (attitude) of an aircraft. This includes controlling the pitch, roll, and yaw angles, as well as the altitude and speed. Attitude control is particularly important for maintaining stability and ensuring that the aircraft responds correctly to pilot inputs and external disturbances. RL can be applied to flight and attitude control to develop controllers that learn optimal control strategies through interaction with the environment. By continuously updating their control policies based on feedback from the aircraft's sensors and performance, RL-based controllers can adapt to changing conditions and uncertainties in the aircraft's dynamics. Implementing RL in flight and attitude control systems requires careful consideration of safety and robustness, as any failure could have severe consequences. Therefore, extensive simulation and testing are essential before deploying RL-based controllers in real-world scenarios. Additionally, integrating RL with existing control architectures and ensuring compatibility with aviation standards and regulations are important challenges to address.
    \item Fault tolerant controller: it ensures the safety and reliability of aircraft, especially in the presence of component failures or unexpected disturbances. These systems are designed to detect faults in the aircraft's control surfaces, engines, or other critical systems and then reconfigure the control strategy to maintain stable flight and safe operation. It can be broadly classified into two categories. \textit{Passive fault-tolerant control}: these systems are designed with inherent robustness to handle faults without the need for detection and reconfiguration. They typically use redundant components and conservative control strategies to ensure stability under a range of fault conditions. \textit{Active fault-tolerant control}: These systems actively detect and isolate faults, and then reconfigure the control strategy to compensate for the fault. This can involve switching to backup systems, adjusting control laws, or using alternative control surfaces. RL can play a significant role in enhancing active fault-tolerant control systems by enabling them to learn and adapt to faults in real time. Algorithms can learn to recognize patterns in sensor data that indicate the onset of a fault. By continuously updating their understanding of normal and faulty conditions, RL-based systems can improve their accuracy in detecting and diagnosing faults. Also, once a fault is detected, an RL agent can learn to select the optimal control strategy to maintain a safe flight. This might involve choosing which backup systems to activate or how to redistribute control authority among the remaining functional components.
    \item Aircraft flight planning: it involves determining the optimal route, altitude, and speed for a flight to ensure safety, efficiency, and compliance with regulations. It is a complex task that takes into account various factors such as weather conditions, airspace restrictions, fuel consumption, and air traffic control requirements. Traditional flight planning relies on predefined algorithms and models to calculate the best flight path. However, these methods may not always be able to adapt to real-time changes or unexpected events, such as sudden changes in weather or airspace closures. RL can address these challenges by learning from experience and continuously updating the flight plan based on new information. Reinforcement learning's role can be placed into different aspects such as dynamic route optimization, fuel-efficient flight planning, adaptive er-routing, and integration with air traffic control. 
    \item Airline maintenance: it ensures that aircraft are safe, reliable, and available for service. It involves regular inspections, repairs, and overhauls of various aircraft components. Traditional maintenance strategies often rely on fixed schedules or reactive approaches, which may not be optimal in terms of cost, efficiency, or aircraft availability. Effective airline maintenance requires balancing several factors. \textit{Safety}: Ensuring that all aircraft systems and components meet stringent safety standards. \textit{Reliability}: Minimizing unexpected breakdowns and delays. \textit{Cost}: Managing maintenance costs while maintaining high safety and reliability standards. \textit{Availability}: Maximizing the time aircraft are available for service and minimizing downtime. RL methods can be used to enhance the quality of the following sides of airline maintenance. \textit{Predictive maintenance}: RL can be used to develop predictive maintenance models that learn from historical data and sensor readings to predict when a component is likely to fail. This allows airlines to perform maintenance proactively, reducing unexpected failures and downtime. \textit{Optimal scheduling}: RL algorithms can learn to schedule maintenance activities optimally, considering factors such as aircraft usage patterns, maintenance resource availability, and the cost of downtime. This can help airlines minimize maintenance costs while ensuring high levels of safety and reliability. \textit{Resource allocation}: RL can assist in the dynamic allocation of maintenance resources, such as personnel and equipment, based on real-time needs and priorities. This can improve the efficiency of maintenance operations. \textit{Spare parts inventory management}: RL can be used to optimize the inventory levels of spare parts, balancing the cost of holding inventory with the risk of stock-outs that could lead to maintenance delays.
    \end{itemize}
Figure~\ref{Taxonomy} illustrates the taxonomy of using RL methods in different aviation applications. In the following sections, we try to summarize the utilization of the RL algorithms in these selected applications. To the best of the authors' knowledge, this survey paper is the first study that reviews the RL methods in aviation.   

Also, it is worth mentioning that the simulation environments are publicly available. In the field of aviation, there are several public benchmarks and simulation environments available for training and testing various models and algorithms. These resources are crucial for researchers and practitioners to develop, evaluate, and compare their approaches. Here's an overview of some of the notable ones.

\subsection{Public Benchmarks and Datasets}
\begin{enumerate}
    \item \textit{Airline On-Time Performance Data }(Bureau of Transportation Statistics): This dataset contains information on flight arrival and departure details for commercial flights within the United States, sourced from the U.S. Department of Transportation's Bureau of Transportation Statistics. It is widely used for research in airline operation optimization, delay prediction, and network analysis \cite{OTDelay}.
    \item \textit{ACAS Xu Dataset}: The Airborne Collision Avoidance System X (ACAS X) is a family of collision avoidance systems developed by NASA. The ACAS Xu dataset is used for developing and evaluating collision avoidance models for unmanned aircraft \cite{ACASX}.
    \item \textit{EUROCONTROL's DDR2 Dataset}: The Demand Data Repository 2 (DDR2) by EUROCONTROL provides air traffic management (ATM) related data, including flight plans, sector configurations, and traffic counts, which are useful for ATM research and simulations \cite{eurocontrol}.
    \item \textit{Flight Quest Dataset}: As part of the GE Flight Quest challenge on Kaggle, this dataset contains flight data aimed at improving flight efficiency, including information on routes, weather, and airspace constraints.
\end{enumerate} 
\subsection{Simulation Environments}
\begin{enumerate}
\item \textit{BlueSky ATC Simulator}: BlueSky is an open-source air traffic control simulator that is used for research and education in air traffic management. It allows for the simulation of airspace, aircraft, and ATC operations \cite{hoekstra2016bluesky}.
\item \textit{JSBSim Flight Dynamics Model}: JSBSim is an open-source flight dynamics model that can be used for flight control analysis, aircraft design, and flight simulation. It provides a flexible framework for modeling the dynamics of fixed-wing and rotary-wing aircraft \cite{berndt2004jsbsim}.
\item \textit{FlightGear Flight Simulator}: FlightGear is a free, open-source flight simulator that can be used for research, education, and pilot training. It offers a realistic flight dynamics model and supports a wide range of aircraft \cite{flightgear}.
\item \textit{OpenAI Gym}: While not specific to aviation, OpenAI Gym provides a standardized interface for reinforcement learning environments, including some that can be adapted for aviation-related tasks, such as control and navigation.
\end{enumerate} 
These resources provide valuable data and simulation capabilities for various applications in aviation, including flight dynamics, air traffic control, route optimization, and collision avoidance. Researchers and developers can leverage these tools to advance the state-of-the-art in aviation technology and safety.

 \subsection{Collision Avoidance and Separation Assurance}

Air traffic control (ATC) plays a crucial role as it is responsible for maintaining flight safety and efficiency. Collision avoidance is the last layer of defense against mid-air collision. Air traffic controllers must maintain a safe separation distance between any two aircraft at all times. This function is called conflict resolution or separation assurance \cite{wang2022review}. An early adaptation of an in-air collision avoidance system was the Traffic Alert and Collision Avoidance System (TCAS) \cite{harman1989tcas} and more recently Next-Generation Airborne Collision Avoidance System (ACAS-X) \cite{jeannin2015formal,kochenderfer2012next}. The latter was built upon TCAS, introducing a partially observable Markov decision process (POMDP) for the problem formulation. It provides audible and visual warnings to pilots by evaluating the time to closest approach, to determine if a collision is likely to occur. Many studies have been recently conducted on RL-based collision avoidance and separation assurance, which a selection is presented in Table~\ref{table:Collision}.
\begin{landscape}
\begin{table*}[t!]

	\centering
	\caption{Selection from the literature on RL in collision avoidance. State/Action space (S/A Space) can be continuous (C), discrete (D), or mixed (M). }
 \scriptsize{
	\begin{tabular}{@{}ccccl@{}}
		\hline
		Reference &
		S/A Space &
		Algorithm &
		Policy Class &
		Key Features \\ \hline
		Wulfe  \cite{wulfe2017uav} &
		D/D &
		Double DQN &
		$\epsilon$-greedy &
		Prioritized sampling, regularization, and discretization of dynamics. \\
		Hermans   \cite{hermans2021towards} &
		M/D &
		DQN &
		&
		Deep Q-learning from Demonstrations, Reward Decomposition. \\
		\begin{tabular}[c]{@{}l@{}}Brittain and \\ Wei   \cite{brittain2019autonomous, brittain2020deep, brittain2021one} \end{tabular}
		  &
		M/D &
		\begin{tabular}[c]{@{}c@{}}PPO, Attention network, \\  and LSTM \end{tabular} &
		ANN &
		 \begin{tabular}[c]{@{}l@{}}1. Adopting a multi-agent framework to handle collision avoidance. \\2. Using LSTM to enhance the performance of PPO. \end{tabular} \\
		Guo et al.   \cite{guo2021safety} &
		M/D &
		PPO, Dropout &
		ANN &
		\begin{tabular}[c]{@{}l@{}}Using Monte-Carlo Dropout and data augmentation \\  to improve the safety in unseen environments. \end{tabular}  \\
		Hu et al.   \cite{hu2021obstacle} &
		C/C &
		PPO &
		ANN &
		\begin{tabular}[c]{@{}l@{}}Developing continuous control for unmanned aircraft system.\end{tabular}\\
		Pham et al.   \cite{pham2019reinforcement} &
		C/C &
		DPG &
		ANN &
		Developing an air traffic scenario simulator. \\
		Wang et al.   \cite{wang2019deep} &
		C/C &
		K-Control Actor-Critic &
		ANN &
		Two-dimensional continuous action selection. \\ 
				Li \textit{et al.}   \cite{li2019optimizing} &
		C/C &
		DPG &
		ANN &
		\begin{tabular}[c]{@{}l@{}}1. Building on ACAS to provide corrections for dense airspace. \\ 2. Handling dense airspace.  \end{tabular}   \\
		Bertram \textit{et al.}   \cite{bertram2020distributed} &
		C/C &
		Fast MDP &
		- &
		\begin{tabular}[c]{@{}l@{}}1. Introducing solution for high-density UAM airspace. \\ 2. Using an MDP-based trajectory planner \\  to avoid cooperative and non-cooperative aircraft.\\ 3. Adopting a multi-agent system to handle large numbers of aircraft.\end{tabular}\\
		Herman  \cite{harman1989tcas} &
		D/D & DDQN & $\epsilon$-greedy &
		\begin{tabular}[c]{@{}l@{}}1. Introducing an onboard collision avoidance tool for pilots.\\ 2. Interrogating an airspace with rule-based logic. \end{tabular}  \\
		Jeannin \textit{et al.}  \cite{jeannin2015formal} & M/D & DQN & -	&
		Formal verification of ACAS-X. \\
		Kochenderfer  \textit{et al.}   \cite{kochenderfer2012next} &
		M/D & PPO, Dropout &
		ANN &
		\begin{tabular}[c]{@{}l@{}}Building on TCAS using a numeric lookup \\ optimized to a probabilistic model. \end{tabular}  \\
		
		\hline
	\end{tabular}
 }
\label{table:Collision}
\end{table*}
\end{landscape}

A MDP collision avoidance in free-flight airspace was introduced in \cite{bertram2020distributed}. In a 3D environment with both cooperative (aircraft actively trying to avoid others) and non-cooperative aircraft (those not concerned with collision avoidance), the MDP formulation in a free flight was able to avoid collision between aircraft. In \cite{li2019optimizing}, a Deep Reinforcement Learning (DRL) method was implemented as an optimization to a collision avoidance problem.

Showing beyond human-level performance in many challenging problems, the collision avoidance problem of unmanned aerial vehicles (UAV) has been solved by implementing a DQN algorithm \cite{wulfe2017uav}. Deep Q-Learning from demonstrations (DQfD) and reward decomposition were implemented to provide interpretable aircraft collision avoidance solutions in \cite{hermans2021towards}. A DQN technique was also applied for the collision avoidance of UAVs \cite{wulfe2017uav}, change routes and speeds in NASA's Sector 33 \cite{brittain2018towards}, compute corrections on top of the existing collision avoidance approaches \cite{harman1989tcas,kochenderfer2012next}, and unmanned free flight traffic in dense airspace \cite{li2019optimizing}. A framework using RL and GPS waypoints to avoid collisions was suggested in \cite{jacob2022autonomous}. A double deep Q-network (DDQN) was applied to guide the aircraft through terminal sectors without collision in \cite{xu2021method}.
The approach tackles the cases where traditional collision avoidance methods fail namely in dense airspace, those expected to be occupied by UAVs, and demonstrated the ability to provide reasonable corrections to maintain sufficient safety among aircraft systems. An Intelligent and Safe Urban Air Mobility (ISUAM) system, which leverages DRL models to execute tactical deviation maneuvers within urban air mobility environments was introduced in \cite{garcia2023isuam}, which the Dueling DQN showed the best performance in terms of satisfying the conflict resolution.

PPO methods are widely used in aircraft collision avoidance and have shown promising success. The problem of collision avoidance in structured airspace using PPO networks \cite{brittain2019autonomous} was addressed using a Long Short-term Memory (LSTM) network \cite{brittain2020deep}, and attention networks \cite{brittain2021one} to handle a variable number of aircraft. While these algorithms show high performance in the training environment, a slight change in the evaluation environment can decrease the performance of these PPO models. A safety module based on Monte-Carlo Dropout \cite{gal2016dropout} and execution-time data augmentation was proposed to solve the collision avoidance problem in environments, which are different from the training environments \cite{guo2021safety}. A PPO network was proposed for unmanned aircraft to provide safe and efficient computational guidance of operations \cite{hu2020uas} and guided UAV in continuous state and action spaces to avoid collision with obstacles \cite{hu2021obstacle}. A message-passing network \cite{dalmauair} was introduced to support collision avoidance. 
 
A prior physical information of airplanes was injected to build a physics-informed DRL algorithm for aircraft collision avoidance \cite{zhao2021physics}. A reward engineering approach was proposed in \cite{panoutsakopoulos2022towards} to support the PPO network to solve the collision avoidance problem in a 2D airspace.

Several studies have applied DDPG \cite{lillicrap2015continuous} to aircraft collision avoidance problems. A DRL method was applied to resolve the conflict between two aircraft with continuous action space in the presence of uncertainty based on DPG in \cite{pham2019reinforcement}. Also, an intelligent interactive conflict solver was used to acquire ATCs' preferences and an RL agent to suggest conflict resolutions capturing those preferences \cite{tran2019intelligent}. Later, the DDPG algorithm dealt with air sectors with increased traffic \cite{tran2020interactive}. A proper heading angle was obtained by the DDPG algorithm before the aircraft reached the boundary of the sector to avoid collisions \cite{wen2019application}. DDPG method was also proposed to mitigate collisions in high-density scenarios and uncertainties in \cite{pham2019machine}. A mixed approach, which combines the traditional geometric resolution and the DDPG model, was proposed to avoid the conflicts \cite{ribeiro2020determining}. Multi-agent deep deterministic policy gradient (MADDPG) was applied to pair-wisely solve the collisions between two aircraft \cite{isufaj2021towards}. Another MADDPG-based conflict resolution method reduced the workloads of ATC and pilots in operation \cite{lai2021multi}.

The actor-critic algorithms are also popular in this application. $K$-control actor-critic algorithm was proposed to detect conflict and resolution with a 2D continuous action space in \cite{wang2019deep}. A policy function returns a probability distribution over the actions that the agent can take based on the given state. A graph-based network for ATC in 3D unstructured airspace was built in \cite{mollinga2020autonomous} to manage the airspace by avoiding potential collisions and conflicts. A multi-layer RL model was proposed to guide an aircraft in a multi-dimensional goal problem \cite{zu2021multi}. Also, an LSTM network and an actor-critic model were used to avoid collisions for fixed-wing UAVs \cite{zhao2021reinforcement}. A recent study aimed to enhance autonomous self-separation in Advanced Air Mobility (AAM) corridors by implementing speed and vertical maneuvers \cite{alvarez2023towards, brittain2024improving}. The research utilized a sample-efficient, off-policy soft actor-critic algorithm to guarantee both safe and efficient aircraft separation.

Besides the popular models, other RL methods were also implemented for collision avoidance. A message-passing-based decentralized computational guidance algorithm was proposed in \cite{yang2020scalable}, which used a multi-agent Monte Carlo tree search (MCTS) \cite{chaslot2008monte} formulation. It was also able to prevent loss of separation (LOS) for UAVs in an urban air mobility (UAM) setting.
A highly efficient MDP-based decentralized algorithm was established to prevent conflict with cooperative and non-cooperative UAVs in the free flight airspace in \cite{bertram2020distributed}. The
MuZero algorithm \cite{schrittwieser2020mastering} was proposed to mitigate a collision in \cite{yilmaz2021deep}.
Difference rewards tool was applied in \cite{singh2021approximate} and a graph convolutional reinforcement learning algorithm solved the multi-UAV conflict resolution problem \cite{isufaj2022multi}. An MARL approach was proposed to mange high-density AAM structured airspace \cite{deniz2024reinforcement}. 

Incorporating a DRL model to learn a collision avoidance strategy while training an NN simultaneously could reduce the learning time and execute a more accurate model due to removing the discretization problem \cite{julian2016policy}. Though DRL has shown great success in aircraft separation assurance, there are still a lot of unsolved problems. These problems create crucial obstacles to building DRL models in this safety-critical application in the real world. One major problem is validation. DRL models for aircraft separation have deep structures and complex input states. The complex architecture makes it difficult to verify the properties of DRL models using traditional formal methods. Current work with formal methods can only validate very simple properties with shallow DRL models. The lack of validation limits the trustworthiness of these DRL models and their use in real-world applications.

Another important question is the gap between simulation and reality. The RL model for aircraft separation assurance is trained with simulators because real-world training is too expensive considering the potential damage. However, it is not possible to have a simulation mimic reality perfectly. The distribution shift between the simulation and reality may constrain the learning performance of the RL models.

Besides these two issues, RL for aircraft separation assurance also faces the problems of general RL models. For example, an RL model currently has a low sampling efficiency, which highly restricts the training speed. Also, the RL model for the separation assurance model works as a black-box. It cannot provide explainable decision-making in this process.

 \subsection{Air Traffic Flow Management}

Air traffic management is an encompassing term for a system that directly affects or is used to decide air traffic movements. The overarching aim of these systems is to reduce delay while maintaining the operational safety of the airspace. Generally, air traffic flow and capacity management are part of a common air traffic service (ATS) and interface with either pilots directly or through ATC. These systems can be considered through two classifications; systems for unmanned traffic management (UTM) and UAS operations, and those for more conventional operations.
\begin{landscape}
\begin{table*}[t!]
	\centering
	\caption{Selection from the literature on RL in air traffic flow management. }
 \scriptsize{
	\begin{tabular}{@{}ccccl@{}}
		\hline
		Reference &
		S/A Space &
		Algorithm &
		Policy Class &
		Key Features \\ \hline
		
		Cruciol \textit{et al.}  \cite{CRUCIOL2013141} &
		C/C &
		Agent-based MARL &
		Q-learning &
		\begin{tabular}[c]{@{}l@{}}1. Reward function considering safety  \\and fairness in Ground Holding Problem (GHP).\\ 2. Reward function considering safety  \\and separation in Air Holding Problem.\end{tabular} \\ 
		
		Xu \textit{et al.}  \cite{xu2020synchronised} &
		C/C & \begin{tabular}[c]{@{}c@{}} K-Control  and\\ Actor-Critic \end{tabular} &
		ANN &
		\begin{tabular}[c]{@{}l@{}}1. A collaborative approach to DCB. \\ 2. Relax constraints of airspace configurations \\ to optimize airspace utilization. \end{tabular} \\ 
		
		Huang \textit{et al.}  \cite{9594397} &
		C/C &
		\begin{tabular}[c]{@{}c@{}} MAA3C  and\\ LSTM \end{tabular} &
		ANN &
		\begin{tabular}[c]{@{}l@{}}Unsupervised and supervised frameworks for ATFM\end{tabular} \\ 
		
		Xi \textit{et al.}  \cite{xie2021reinforcement} &
		C/C &
		DQN &
		ANN &
		\begin{tabular}[c]{@{}l@{}}1. Flow management for UAM airspace. \\ 2. Using a DQN with genetic algorithm to solve DCB problem. \end{tabular} \\
		Tang and Xu  \cite{tang2021multi} &
		C/C &
		\begin{tabular}[c]{@{}c@{}} K-Control  and \\ Actor-Critic \end{tabular} &
		ANN &
		\begin{tabular}[c]{@{}l@{}}1. Rule-based time-step environment to mimic the DCB process. \\ 2. MARL framework to address credit assignment problem.\end{tabular} \\
		\begin{tabular}[c]{@{}l@{}}Spatharis \textit{et al.} \\ \cite{kravaris2019resolving, spatharis2021hierarchical}\end{tabular} &
		C/C &
		Hierarchical RL &
		$\epsilon$-greedy &
		\begin{tabular}[c]{@{}l@{}}1. Hierarchical approach partitions task  \\into hierarchies of states and actions.\\ 2. Hierarchical methods improve on DCB in the pre-tactical stage. \\
        3. Agent-based MARL. \end{tabular} \\ 
		
		Duong \textit{et al.}  \cite{duong2019decentralizing} &
		C/C &
		Block-chain based RL &
		$\epsilon$-greedy &
		\begin{tabular}[c]{@{}l@{}}Introduces decentralised a blockchain-based RL agent.\end{tabular} \\ 
    
    Chen \textit{et al.} \cite{chen2021demand} & C/C & DDQN, experience replay & Adaptive $\epsilon$-greedy & \begin{tabular}[c]{@{}l@{}} 1. Comparison with the actual method used in operations \\  2. Decentralized Training with Decentralized Execution \end{tabular} \\

		\hline
	\end{tabular}
 }
	\label{table:trafiicmanagement}
\end{table*}
\end{landscape}

Air Traffic Flow and Management (ATFM) is a subset of traffic management that focuses on ensuring the available airspace capacity is used efficiently. The capacity can be influenced by the sector's geometry: size, shape, or altitudes as well as stochastic variables like wind, weather, and emergencies, or more constant variables like airport capacity and throughput. In choosing the role of an autonomous system, existing workflows must be observed with human actors to identify what could maximize performance. Using a MARL approach \cite{schrittwieser2020mastering} presents two reward functions for both the Ground Holding Problem (GHP) and Air Holding Problem (AHP). The approach focuses on optimizing six objectives: minimize delays in the ATC sector; minimize delays in the terminal sector; minimize financial cost to airlines; improve fairness among airlines; reduce impact in ATC sectors by avoiding unnecessary actions; and improve safety in the ATFM. Based on these factors, the reward functions, when tested in Brazilian airspace, maintained safety standards when aiding air traffic controller decision-making and also improved efficiency and fairness among aircraft.

Demand capacity balancing (DCB) is a predictive method to ensure the efficient operation of airspace or ground operations. A collaborative approach was introduced to DCB utilizing: assigning delays, allowing alternative trajectories, using fixed airspace sectorisation, or adjusting airspace sectorisation to efficiently manage airspace \cite{xu2020synchronised}. Unlike other solutions, synchronized collaborative-demand capacity balancing (SC-DCB) seeks to relax the constraints of airspace configurations, with the outcome demonstrating a reduction in active sectors resulting in better utilization of the active ones. Further modeling the DCB problem as a POMDP, \cite{9594397} proposes a multi-agent asynchronous advantage actor-critic (MAA3C) network to manage delaying aircraft based on an unsupervised training approach. A combination of DCB for strategic conflict management and RL for tactical separation was proposed in \cite{chen2024integrated}. A better tactical safety separation performance was achieved by using DCB to precondition traffic to proper density levels. Also, a recent study showed the efficiency of RL methods by changing the traffic density in the training process \cite{groot2024analysis}.

By improving the computational capabilities, flights have been considered as agents, and MARL methods \cite{spatharis2018multiagent} were proposed to solve the capacity problems. Various algorithms were also studied: independent learners, edge-MARL, and agent-based-MARL, based on Q-learning techniques. Inspired by supervised learning, multiple supervised-MARL frameworks built on PPO were suggested \cite{tang2021multi}, where the agents representing the flights have three actions: hold their departure, take-off, or cooperate. This study indicated that adding supervisors can help improve search and generalization abilities. DQN and decentralized training and decentralized execution (DTDE) combined with replay experience \cite{chen2021demand} were also used to solve the DCB problem.

In recent work \cite{xie2021reinforcement}, RL techniques have been utilized to examine their efficiency in UAM flow management, using a state space consisting of data retrieved from aircraft, weather, airspace capacity, and traffic density surveillance, and training data constructed through a Post-Hoc system. Multi-agent approaches in flow management also emerged \cite{duong2019decentralizing, tang2021multi, spatharis2021hierarchical} to demonstrate that a MARL approach can successfully resolve hot spots in dense traffic areas by taking holding, departure, or cooperation actions, resulting in an overall reduction in delay.

%%% Related works on taxi-out and ground delay prediction %%% 

%Predicting the correct arrival time of aircraft plays a vital role in a nation's aviation industry. The taxi-out time is the average time an aircraft spends between the departure from a gate and take-off. One of the most critical uncertainties in this process is taxi-out delays. Taxi-out delay is also a significant portion of flight delays which causes customers dissatisfaction. It can account for 60\% of the total delay at significant airports; accurate prediction could help improve ground efficiency. 
%The taxi-out time prediction problem was considered as an MDP problem \cite{lee2016taxi}, and the Bellman optimality equation was used to solve this reinforcement learning problem. 
% An RL-based method was proposed to predict the taxi-out time for a Tampa International Airport case study \cite{balakrishna2010accuracy}. The utilized method, on average, with 93.7\% probability, predicted the mean taxi-out time for any quarter with a 1.5-min error. It also predicted 81\% of the taxi-out time for individual flights with a two-min error. The promising results of the proposed method indicated that the RL estimator has an excellent potential to capture the dynamics at even challenging airports such as the New York airports.

Ground delay programs (GDP) deal with an excessive number of flights reaching an airport serving as another air traffic flow management mechanism. Airports' ability to handle arrivals may be adversely affected by weather conditions. Hot spots were solved using GDP, in which flight departures are delayed, to shift the whole trajectory \cite{kravaris2019resolving}. The results showed that collaborative methods yield better results.
To reduce the search space, a hierarchical MARL scheme was proposed to solve the DCB problem with GDP \cite{spatharis2021hierarchical}, thus allowing the abstraction of time and state-action.

Issuing terminal traffic management initiatives (TMIs) is a technique for reducing the number of incoming aircraft to an airport for a short period. One type of this technique is the ground delay program.
A data-driven approach based on a multi-armed bandit framework was proposed for suggesting TMI actions \cite{estes2017data}. This would be beneficial for human decision-makers to evaluate whether a suggested solution is reasonable or not. The suggestions were based on historical data of forecasted and observed demand and capacity, chosen TMI actions, and observed performance. The results showed that almost all proposed algorithms slightly outperform the historical actions. \cite{james2021tmi} proposed four methods for recommending strategic TMI parameters during uncertain weather conditions. The first two methods were based on random exploration, while the others were using an $\epsilon$-greedy approach and a Softmax algorithm. The fast-time simulation results demonstrated the strong performance of the two latter methods relative to the others, and their potential to help with dealing with weather uncertainty.

A comparison between behavioral cloning (BC) and inverse reinforcement learning (IRL) in predicting hourly expert GDP implementation actions was made in \cite{bloem2015ground}. Historical data was used to predict GDP decisions on San Francisco and Newark international airports. The IRL method was proposed to reduce the complexity by only exploring the states in the data. The results demonstrated that BC has a more robust predictive performance than the IRL GDP-implemented models. The experiments also suggested that neither the BC nor the IRL models predict the relatively infrequent GDP initialization or cancellation events well, unlike Q-learning, which tends to provide accurate predicted times \cite{george2015reinforcement}. Better prediction of taxi-out times will improve taxiing management, which can benefit trajectory planning by using GDP to reduce congestion.

Runway Configuration Management (RCM) plays a crucial role in optimizing runway usage, considering factors such as traffic flow and weather conditions. This complex facet of air traffic management is challenging due to its reliance on fluctuating operational and environmental conditions. 
A Runway Configuration Assistance (RCA) decision-support tool was developed by employing offline model-free reinforcement learning \cite{memarzadeh2023airport, nethi2024optimization}. An innovative integration of predictive data from the Localized Aviation Model Output Statistics Program (LAMP) and Terminal Area Forecast (TAF) were introduced, where significantly improving the tool's precision and its responsiveness to rapid changes in wind conditions.

With airspace becoming denser due to higher traffic and the introduction of emerging UAS/UTM technologies, traffic management solutions will be needed to demonstrate the ability of adaptation to accommodate not only higher volumes and densities of air traffic but also any new requirements imposed by this new classification of air traffic. Additionally, the safety and capacity of these systems will require formal verification and standardized validation, moving the field of RL in ATM away from the laboratory and being ready to be accepted by official bodies. Finally, there are still many unknowns about how the UTM/UAS airspace will be constructed, adding a further layer of complexity to solution design; new systems should entertain this notion and provide flexibility while the airspace is still being defined. 

\subsection{Airline Revenue Management}
In 1970s, there was limited control over ticket pricing and network scheduling. If one airline company wanted to increase its fare, permission from a federal agency called the Civil Aeronautics Board (CAB) was needed. The pricing regulation at that time always led to a higher fare. Airline deregulation happened in 1979, which allowed companies to schedule and price freely. Consequently, airline revenue management (ARM) came out as a business practice to set prices when there is perishable inventory. The ARM is an airline company's strategy to maximize revenue by optimizing ticket prices and product availability. The classic ARM problem could be divided into two types, quantity-based and price-based revenue management (RM) \cite{talluri2004theory}. 
\begin{landscape}
\begin{table*}
	\centering
	\caption{Selection from the literature on RL in airline revenue management.}
	\label{table:revenue}  
 \scriptsize{
	\begin{tabular}{@{}ccccl@{}}
		\hline
		Reference  & Problem Type & Algorithm & Policy Class& Key Features \\ \hline
		Gosavi \textit{et al.} \cite{gosavii2002reinforcement} &Single leg&  Q-learning & $\epsilon$-greedy & 
		\begin{tabular}[c]{@{}l@{}}	Infinite time horizon under the average reward\\ optimizing criterion. \end{tabular} \\
		Lawhead and Gosavi \cite{lawhead2019bounded}  &Single leg & \begin{tabular}[c]{@{}c@{}}	Bounded \\  Actor-Critic \end{tabular} 	    &$\epsilon$-greedy&
		\begin{tabular}[c]{@{}l@{}}		Test two types of reward: discounted reward MDP  \\and the average reward SMDP. \end{tabular}\\
		Bondoux \textit{et al.} \cite{bondoux2020reinforcement}& Single leg& DQN  &$\epsilon$-greedy&Comparison between DQL and RMS \\
		Shihab and Wei \cite{shihab2021deep}&Single leg& DQN & $\epsilon$-greedy&
		\begin{tabular}[c]{@{}l@{}}	Considering both cancellation  \\ and overbooking in the environment. \end{tabular}\\
		Wang \textit{et al.} \cite{wang2021solving}& Single leg & DQN Actor-Critic &$\epsilon$-greedy&
		\begin{tabular}[c]{@{}l@{}}	Combining quantity-based RM  and price-based RM together. \end{tabular}\\
		Alamdari and Savard \cite{alamdari2021deep}& Single leg \& Network& DQN &AGen&
		\begin{tabular}[c]{@{}l@{}}	Greedily generate a set of ``effective'' actions to replace\\the original action space. \end{tabular} \\ \hline
\end{tabular}
}
\end{table*}
\end{landscape}

Quantity-based RM works on a predefined n-class fare structure and determines how many tickets are protected for each fare class. Also, it focuses on the capacity control of single and network flight legs. As a representative of the quantity-based RM, the expected marginal seat revenue (EMSR) models \cite{belobaba1987air} are wildly used in the modern airline industry. The price-based RM focuses more on the dynamic pricing situation.

Traditional and widely used approaches for ARM systems are model-based and data-driven, which heavily depend on the accuracy of forecasting data such as passenger arrival distribution, willingness to pay (WTP), and cancellation rate. Recently, researchers have been considering applying model-free learning-based methods on ARM, such as optimal control theory or RL. A research direction of using RL in ARM started in 2002 \cite{gosavii2002reinforcement}, where the $\lambda$-smart algorithm was designed to cast the single-leg ARM problem as a semi-Markov decision problem (SMDP) over an infinite time horizon under the average reward optimizing criterion. Later, a bounded actor-critic approach was applied on the same problem \cite{lawhead2019bounded}. Both studies claimed that the model's performance was better than the EMSR model. A DRL model on ARM has been introduced to integrate the domain knowledge with a DNN trained on graphical processing units (GPUs) \cite{bondoux2020reinforcement}. A DRL model was also applied to the inventory control problem, using DQN and considering both cancellation and overbooking in their environment \cite{shihab2021deep}. Some other improvements to DRL models have also appeared in recent years. For example, an ARM problem was studied by combining quantity-based RM, and price-based RM \cite{wang2021solving}, while the DRL was applied to both the single leg and network leg problems \cite{alamdari2021deep}.

The previous learning-based approaches consider the game between passengers and airline companies. However, there is limited work regarding the competitive pricing process among different airline companies. We believe it will be an exciting topic with the development of MARL.

 \subsection{Aircraft Flight and Attitude Control}

Attitude control of an aircraft can be challenging due to the system's nonlinearities, uncertainties, and noises acting upon the system, which are intrinsically present in the environment. Recently, researchers have aimed to develop advanced controllers based on RL algorithms. A selection of RL methods in attitude control applications is presented in Table~\ref{table:AttitudeControl}. This section focuses on controllers based on RL algorithms. A comprehensive survey on NN-based flight control systems was done in \cite{Emami_2022}.

These proposed controllers have been used in target tracking \cite{li2022adaptive,li2021maneuvering}, single/multi-agent obstacle avoidance \cite{li2021maneuvering,zhao2020research}, vision-based landing \cite{lee2018vision}, stabilization \cite{xian2021robust,zhen2020deep,huang2019attitude,huang2020model}, visual servoing \cite{shi2016decoupled}, and flat spin recovery \cite{kim2017reinforcement}.
In \cite{huang2019attitude}, it was shown that training a controller directly by RL, based on a nonlinear or unknown model, is feasible. The performance of the controllers based on different RL algorithms was also compared in \cite{zuo2019flight}. The results showed that a DQN is more suitable for discrete tasks than policy gradient or DDPG, whereas DDPG was shown to perform better in more complex tasks. Also, a DQN method was used to design attitude control systems for aircraft \cite{huang2019attitude,zuo2019flight}. In addition, the DDPG-based controllers were established in \cite{li2021maneuvering,zuo2019flight,wang2020attitude, zhang2020autonomous,al2019machine}. An improved DDPG method was combined with transfer learning and a control system was developed to perform autonomous maneuvering target tracking \cite{li2021maneuvering}. A DDPG-based controller was also studied, guiding a UAV to a fixed position in a horizontal plane from any position, and attitude \cite{zhang2020autonomous}.

Other studies have been conducted using PPO methods \cite{zhao2020research,zhen2020deep,bohn2019deep}. An improved MARL algorithm was developed, named multi-agent joint proximal policy optimization (MAJPPO), to perform formation and obstacle avoidance. The controller has used a moving averaging method to make each agent obtain a centralized state value function \cite{zhao2020research}. By performing the experimental comparison, it was shown that the MAJPPO algorithm could better deal with partially observable environments. A PPO-based controller was designed for stabilizing a fixed-wing UAV \cite{zhen2020deep}. It was shown that the RL controller could stabilize the system in the presence of disturbances in the environment more precisely compared to a PID controller. 

Since RL has achieved significant progress in attitude control, it has been considered a promising approach for designing optimal and robust controllers. However, there are still some challenges that should be addressed. The gap between simulations and natural environments was experimentally demonstrated \cite{wada2021unmanned}, which required a new training approach. A controller learned to adapt to the difference between training models and real environments. Exploration and exploitation balance is another dilemma in RL. A normal distribution noise for exploring the environment was used at the start of the training process \cite{wang2020attitude}. It also proposed using Uhlenbeck-Ornstein stochastic noise for future works. 
\begin{landscape}
\begin{table*}[t!]
	\centering
	\caption{Selection from literature on RL in attitude control.}
	\label{table:AttitudeControl}
 \scriptsize{
	\begin{tabular}{@{}ccccl@{}}
		\hline
		Reference    & S/A Space & Algorithms   & Policy Class       & Key Features  \\ \hline
	Li \textit{et al.} \cite{li2022adaptive}      & C/C   & Actor-Critic    & ANN  & \begin{tabular}[c]{@{}l@{}}1. Compensating for the actuator fault and system input saturation.\\ 2. Proving system stability by Lyapunov theory.\end{tabular} \\
		Li \textit{et al.} \cite{li2021maneuvering}   & C/C   & \begin{tabular}[c]{@{}c@{}}MMN-DDPG \\ Transfer Learning \end{tabular}   & ANN  &
		\begin{tabular}[c]{@{}l@{}}1. Introducing exploratory noises and parameter-based \\ transfer learning to improve speed and generalization. \\ 2. Performing target tracking and obstacle avoidance \\ precisely in uncertain environments.  \end{tabular} \\
		Zhao \textit{et al.}  \cite{zhao2020research}   & -     & \begin{tabular}[c]{@{}c@{}}Multi-agent joint PPO \\ (MAJPPO) \end{tabular}  & ANN & \begin{tabular}[c]{@{}l@{}}1. Using a moving window averaging of state-valued function  \\ to deal with multi-agent coordination problems.
		\\	2. MAJPPO, a centralized training and distributed execution.
		\end{tabular} \\
		Lee \textit{et al.}   \cite{lee2018vision}        & C/C                & Actor-Critic                   & ANN                & \begin{tabular}[c]{@{}l@{}}Using a simple PID controller for handling attitude and  \\ position of UAV and a DRL algorithm to generate proper commands.\end{tabular} \\
		Xian \textit{et al.}  \cite{xian2021robust}        & C/C                & Actor-Critic                   & ANN                & \begin{tabular}[c]{@{}l@{}}1. Compensating the error of actor-critic network  \\ by a robust nonlinear sliding mode control method.
		\\2. Achieving a better control performance compared to LQR.
		\end{tabular} \\
		Zhen \textit{et al.}   \cite{zhen2020deep}          & -      & PPO      & ANN   & \begin{tabular}[c]{@{}l@{}}Achieving more precise control comparing  \\to a PID controller in the presence of disturbance. \end{tabular} \\                                                                 
		Huang \textit{et al.}  \cite{huang2020model}          & C/C                & Actor-Critic                   & ANN                & \begin{tabular}[c]{@{}l@{}}Introducing an NN approximation to learn  \\ the optimal controller online with no information of model.\end{tabular}                                                                                                                                         \\
		Huang \textit{et al.}  \cite{huang2019attitude}        & C/C                & DDQN                           & ANN / $\epsilon-\text{greedy}$ & \begin{tabular}[c]{@{}l@{}}Proposing model can train  the controller  \\ in time domain directly on nonlinear or unknown model.
		\end{tabular} \\
		Shi \textit{et al.}  \cite{shi2016decoupled}    & D/D                & Q-learning                     & TD                 &  Taking Q-learning for adaptive servoing gain adjustment.  \\
		Kim \textit{et al.}  \cite{kim2017reinforcement}     & D/C                & DQN                            & ANN                & Covering both unusual attitude and stable spin mode recoveries.   \\                                                             
		Zuo \textit{et al.} \cite{zuo2019flight}   & C/C                & DQN, PG ,  DDPG                & ANN / $\epsilon$-greedy & \begin{tabular}[c]{@{}l@{}}
			1. Being more efficient and faster. \\
			2. Handling continuous action space but not efficient enough.
		\end{tabular}  \\
		Wang \textit{et al.} \cite{wang2020attitude}    & C/C                & DDPG                           & ANN                & Using a normal distribution for having better exploration.  \\
		Bohn \textit{et al.} \cite{bohn2019deep}  & C/C                & PPO                            & ANN                & \begin{tabular}[c]{@{}l@{}}1. Converging faster than PID. \\ 2. Generalizing to turbulent wind conditions.\end{tabular}                                                                                                                                         \\
		Wada \textit{et al.} \cite{wada2021unmanned} & C/C    &\begin{tabular}[c]{@{}c@{}}Actor-Critic(A3C)\\
		LSTM \end{tabular}  & ANN       & \begin{tabular}[c]{@{}l@{}}	1. Stability of the NNs in different delays. \\
		2. Experimentally demonstrating the reality and simulation gap. 
		\end{tabular} \\ \hline
	\end{tabular}
 }
\end{table*}
\end{landscape}

 \subsection{Fault Tolerant Controller}
A fault is a change in a system's property or parameters that causes the system to behave differently from its design. In other words, failure is a condition that prevents a system from functioning. A fault-tolerant controller (FTC) is a control strategy that aims to improve the performance of a system operating in degraded performance due to a fault \cite{blanke2006diagnosis}. FTCs are characterized as model-based or data-driven, based on the method used to develop the controllers. Model-based techniques necessitate knowledge of the system's model and parameters to design a fault-tolerant controller.
On the contrary, data-driven approaches learn the FTC directly from system data. The fundamental problem of a model-based FTC approach is that its effectiveness depends on the system model's correctness, which is difficult to establish when system parameters can vary due to faults. Furthermore, complex systems necessitate complicated controllers, which, in turn, impacts the controllers' robustness. On the other hand, data-driven techniques utilize data to design FTC without knowing the system's dynamics. As a result, data-driven methods, particularly RL-based techniques, have recently gained a lot of attention.

Several approaches have been proposed in the literature to solve the FTC controller using RL. Different RL algorithms, including DDPG, TRPO, and PPO, have been used to develop FTC techniques for quadrotor attitude control \cite{koch2019reinforcement}. The results indicated that among the developed RL-based fault-tolerant controllers, the trained PPO-based attitude controller outperformed a fully tuned PID controller in terms of rising time, peak velocities achieved, and total error among the trained set of controllers. A DPG-based technique with an integral compensator was adopted to develop a position-tracking controller for the quadrotor \cite{wang2019deterministic}. The approach employed a two-phased learning scheme, with a simplified model being utilized for offline learning and the learned policy being refined during flight. The results showed that the learned FTC is sufficiently robust to model errors and external disturbances. A DDPG-based fault-tolerant policy for position tracking of quadcopters was proposed in \cite{fei2020learn}. The framework operates so that it runs simultaneously with the model-based controller and only becomes active when the system's behavior changes from the normal operating condition.

One of the significant drawbacks of model-free RL-based FTC methods is that there is no guarantee of convergence. To overcome this problem, a model-based framework for position tracking of octocopters was proposed \cite{9595275}.  Four RL algorithms were proposed, PPO, DDPG, Twin-Delayed DDPG (TD3), and soft actor-critic (SAC). The results showed that PPO is more suitable for a fault-tolerant task. 

\begin{landscape}
\begin{table*}[t!]
	\centering
	\caption{Selection from the literature of RL on fault-tolerant controller.}
	\label{table:fualt}  
 \scriptsize{
	\begin{tabular}{@{}ccccl@{}}
		\hline
 Reference & Problem Type & S/A Space & Algorithm & Key Features \\  
 \hline
 Koch \textit{et al.} \cite{koch2019reinforcement} & Attitude control & C/C &  	DDPG, TRPO, PPO  & Training RL algorithms to perform end-to-end attitude control.  \\
 Wang \textit{et al.} \cite{wang2019deterministic} & Position tracking & C/C & DPG &
 \begin{tabular}[c]{@{}l@{}} Integrating DPG with integral compensator \\ and adopting a two-phased approach. \end{tabular} \\
 Fei \textit{et al.}\cite{fei2020learn}  & Position tracking & C/C & DDPG &  
Running simultaneously with the model-based controller.  \\
  Bhan \textit{et al.}\cite{9595275} & Position tracking & C/C & \begin{tabular}[c]{@{}c@{}}DDPG,TD3 \\ SAC,PPO  \end{tabular}   &\begin{tabular}[c]{@{}l@{}}1. Estimating fault-related parameters using an estimator. \\  2. Training several RL algorithms using the estimated parameters. \end{tabular} \\ \hline
\end{tabular}
}
\end{table*}
\end{landscape}

 \subsection{Aircraft Flight Planning}
Flight and trajectory planning is a well-known aviation problem and is crucial. While airspace users want the most optimal trajectory to minimize a cost function, many constraints, such as ground obstacles, capacity limitations, or environmental threats, make this problem difficult to solve. Several techniques, including rerouting or ground delay are proposed to mitigate traffic congestion in most cases. The ATM domain is essentially based on temporal operations, with a capacity supply and demand model to manage air traffic flows. This operation can lead to capacity imbalances and create hot spots in sectors when capacity (defined as the number of aircraft accepted in a given sector during a given period) is exceeded. The planning of a trajectory or flight of an aircraft can be done in several stages defined in the ATM domain; the strategic phase includes the planning of flights performed between one year and D-7, the pre-tactical phase takes place between D-7 and D-1, and finally, the tactical phase takes place on D-day. An RL planner has shown to be a promising tool to solve pre-flight planning problems in dangerous environments \cite{wickman2021exploring}.  

UAV's versatility in performing tasks ranging from terrain mapping to surveillance and military missions makes this problem a fundamental part of aircraft operations \cite{azar2021drone}. One of many defined missions for UAVs is to fly over ground targets. The theory of POMDP was presented for military use, and nominal belief-state optimization (NBO) was used to find the optimal trajectory considering threats, wind effects, or other agents \cite{ragi2013uav}. Also, an RL approach was proposed to use geometric information from the drone's environment and produce smoother and more feasible trajectories in real-time planning \cite{zhang2015geometric}. The dueling double deep Q-networks (D3QN), DDQN, and DQN methods have been compared in \cite{yan2020towards} to solve the path planning problem for an agent in the context of a dynamic environment where it faces an environmental threat.

An RL method was used to resolve these hot spots with traffic speed regulation \cite{tumer2007distributed}. An agent representing a fix (a 2D point in the sector) can regulate the flows. In addition, a multi-agent asynchronous advantage actor-critic (MAA3C) framework was constructed to resolve airspace hot spots within a proper ground delay \cite{huang2021integrated}.

%%% Table recap of references %%%
\ 

\begin{landscape}
\begin{table*}[t!]
	\centering
	\caption{Selection from the literature on RL in flight planning. }
	\label{table:planning} 
 \scriptsize{
	\begin{tabular}{@{}ccccl@{}}
		\hline
Reference & Problem Type & Algorithms & Policy Class & Key Features \\ \hline
    Ragi \textit{et al.} \cite{ragi2013uav} & UAV path planning & 
    \begin{tabular}[c]{@{}c@{}} POMDP\\ NBO approximation  \end{tabular} & - & \begin{tabular}[c]{@{}l@{}}	1. Dynamical environment. \\  2. Wind effects are taken into account. \end{tabular} \\
    Zhang \textit{et al.} \cite{zhang2015geometric} & UAV path planning & Geometric RL & - & \begin{tabular}[c]{@{}l@{}}	Convergence of calculating  \\  the reward matrix theoretically proven. \end{tabular} \\
    Yan \textit{et al.} \cite{yan2020towards} & UAV path planning & D3QN, DDQN, DQN & $\epsilon$-greedy & \begin{tabular}[c]{@{}l@{}} 1. Stage scenario for simulation. \\ 2. DRL approach and comparison of methods.  \end{tabular} \\ 
    Bertram \textit{et al.} \cite{bertram2021scalable} & Flight plan scheduling & FastMDP & $\epsilon$-greedy & \begin{tabular}[c]{@{}l@{}} 1. Centralized or distributed flight plan scheduling. \\ 2. Parallelization for large-scale scheduling.  \end{tabular} \\ 
\hline

\end{tabular}
}
\end{table*}

\end{landscape}
\normalsize
%%% Little introduction on the subject %%%

%\subsubsection{UAV path planning}

%%% Related works on UAV path planning %%%

%\subsubsection{Flight planning}

%%% Related works on flight planning %%%

All these works aim to reduce hot spots by delaying flights while minimizing average delays and ensuring good distribution. Still, they have not studied other trajectory planning techniques. An RL approach was proposed to select a low-level heuristic to mitigate the air traffic complexity \cite{juntama2022hyperheuristic}. Flight level allocation, staggered departure times, and en route path deviation reduced congestion. In a UAM concept, the pre-departure airspace reservation problem as an MDP was formulated \cite{bertram2021scalable}. The first-in-first-out (FIFO) principle and the fast-MDP algorithm provided a conflict-free trajectory at the strategic stage. The scheduler, allowing both centralized and decentralized flight planning, takes advantage of the computing power and parallelization of GPUs to process a large number of flights. A Learning-to-Dispatch algorithm was proposed to maximize the air capacity under emergencies such as hurricane disasters \cite{zhang2021learning}.

 \subsection{Airline Maintenance}

Maintenance scheduling is the process of planning when and what type of maintenance check should be performed on an aircraft. The maintenance tasks of airlines are usually grouped into four-letter checks (A, B, C, and D). The level of detail in the maintenance check of these groups is different. For example, A- and B-checks are considered light maintenance, and C- and D-check as heavy maintenance and more detailed inspection. Usually, weather conditions and flight disruptions cause deviation from the scheduled plan. These uncertainties make aircraft maintenance scheduling a challenging task.

In addition, amidst the burgeoning competition in the global airline industry, where key market players vie for profitability, there has been a significant focus on devising optimal routes that are maintenance feasible. The primary goal of the Operational Aircraft Maintenance Routing Problem (OAMRP) is to create such optimal and maintenance-viable routes for each aircraft, ensuring compliance with FAA regulations.  The OAMRP through two primary lenses was examined \cite{ruan2021reinforcement}. Initially, a formulation for a network flow-based Integer Linear Programming (ILP) framework was introduced that addresses three critical maintenance constraints concurrently: the maximum allowable flying hours, the number of take-offs permitted between consecutive maintenance checks, and workforce capacity limitations. Subsequently, a novel RL-based algorithm was designed to resolve the OAMRP more swiftly and effectively.

A look-ahead approximate dynamic programming methodology was developed for aircraft maintenance check \cite{deng2022lookahead}. Its schedules minimized the wasted utilization interval between maintenance checks while reducing the need for additional maintenance slots. The methodology was tested with two case studies of maintenance data of an A320 family fleet. The developed method showed significant changes in scheduled maintenance times; it reduced the number of A-checks by 1.9\%, the number of C-check by 9.8\%, and the number of additional slots by 78.3\% over four years.

An RL-based approach was proposed in \cite{hu2021reinforcement} to solve the aircraft's long-term maintenance optimization problem. The proposed method uses information about the aircraft's future mission, repair cost, prognostics and health management, etc., to provide real-time, sequential maintenance decisions. The RL-driven approach outperforms three existing commonly used strategies in adjusting its decision principle based on the diverse data in several simulated maintenance scenarios. The integration of an RL model for Human–AI collaboration in maintenance planning and the visualization of the Condition-Based Maintenance indicators were proposed in \cite{ribeiro2022playful}. Optimal maintenance decision-making in the presence of unexpected events was also developed. 

Also, a recent study introduced an innovative method for predicting unscheduled aircraft maintenance actions by employing DRL techniques and utilizing data from aircraft central maintenance system logs. The proposed algorithm reframes the challenge of predicting rare failures as a sequential decision-making process \cite{dangut2022application}.

 \subsection{Safety and Certification of Reinforcement Learning}

Safety is of utmost importance in safety-critical applications such as aviation systems. Recent promising results in RL have encouraged researchers to apply such techniques to many real-world applications. However, the certification of learning-based approaches, including RL in safety-critical applications, remains an open research question \cite{van2017challenges,baheri2022verification}. Recent surveys provide a comprehensive overview of efforts toward safe RL for safety-critical applications \cite{garcia2015comprehensive}. While there has been a lot of research interest in safe RL, especially in the autonomous driving community \cite{kiran2021deep,baheri2020deep,baheri2022safe}, the safe RL problem is still underexplored in the aviation research community. The application of safe RL in aviation systems has been studied from different angles. For instance, recently, a safe DRL approach was proposed for autonomous airborne collision avoidance systems \cite{panoutsakopoulos2022towards}. From the conflict resolution perspective, soft actor-critic models were used during vertical maneuvers in layered airspace \cite{groot2021improving}. In a similar line of research, a safe deep MARL framework can identify and resolve conflicts between aircraft in a high-density \cite{brittain2019autonomous}.
\\
From the run-time assurance perspective, a run-time safety assurance approach casts the problem as an MDP framework and uses RL to solve it \cite{lazarus2020runtime}. Similarly, the path planning problem was framed as MDP and utilized MCTS for safe and assured path planning \cite{wu2022comparisons}. To guarantee the safety of real-time autonomous flight operations, an MCTS algorithm was proposed along with Gaussian process regression and Bayesian optimization to discretize the continuous action space \cite{wu2022safety}. Furthermore, a reinforcement learning framework predicts and mitigates the potential loss of separation events in congested airspace \cite{hawley2019real}. Recently, a safety verification framework was presented for design-time and run-time assurance of learning-based components in aviation systems \cite{baheri2022verification}.
 \section{Conclusion}
In this paper, after a review of the most common RL techniques and the overall methodology and principles, a survey of the applications of RL in aviation is proposed. Ranging from airline revenue management to aircraft altitude control, the use of RL methods has shown a great interest in the literature in the last decade. Indeed, with the increase in computational power and access to a large source of data, this data-driven approach has become widely studied. Whether it is collision avoidance, traffic management, or other aviation-related problems, these learning-based frameworks show promising results and a variety of algorithms and techniques are often studied for a specific problem. The most advanced techniques such as DRL or DPG are used to deal with critical systems such as collision avoidance or to handle the increase of growing air traffic in traffic management and flight planning. However, differences between the simulated environment and real-world application or its black-box scheme can still be a hindrance to implementation in the aviation industry, constrained by numerous safety measures. The certification of such methods is then a crucial point for these innovative and disruptive applications in aviation and should be one of the focuses of research in this area.

\section*{Acknowledgment}
The paper is disseminated under the sponsorship of the U.S. Department of Transportation in the interest of information exchange. The U.S. Government assumes no liability for the contents or use thereof. The U.S. Government does not endorse products or manufacturers. Trade or manufacturers’ names appear herein solely because they are considered essential to the objective of this paper. The findings and conclusions are those of the authors and do not necessarily represent the views of the funding agency. This document does not constitute FAA policy. This research was supported by the FAA under contract No. 692M15-21-T-00022.

\bibliographystyle{elsarticle-num}
\bibliography{main}

\begin{thebibliography}{100}
\expandafter\ifx\csname url\endcsname\relax
  \def\url#1{\texttt{#1}}\fi
\expandafter\ifx\csname urlprefix\endcsname\relax\def\urlprefix{URL }\fi
\expandafter\ifx\csname href\endcsname\relax
  \def\href#1#2{#2} \def\path#1{#1}\fi

\bibitem{bertsekas2012dynamic}
D.~Bertsekas, Dynamic programming and optimal control: Volume I, Vol.~1, Athena
  scientific, 2012.

\bibitem{ha2018world}
D.~Ha, J.~Schmidhuber, World models, arXiv preprint arXiv:1803.10122 (2018).

\bibitem{feinberg2018model}
V.~Feinberg, A.~Wan, I.~Stoica, M.~I. Jordan, J.~E. Gonzalez, S.~Levine,
  Model-based value estimation for efficient model-free reinforcement learning,
  arXiv preprint arXiv:1803.00101 (2018).

\bibitem{racaniere2017imagination}
S.~Racani{\`e}re, T.~Weber, D.~Reichert, L.~Buesing, A.~Guez,
  D.~Jimenez~Rezende, A.~Puigdom{\`e}nech~Badia, O.~Vinyals, N.~Heess, Y.~Li,
  et~al., Imagination-augmented agents for deep reinforcement learning,
  Advances in neural information processing systems 30 (2017).

\bibitem{nagabandi2018neural}
A.~Nagabandi, G.~Kahn, R.~S. Fearing, S.~Levine, Neural network dynamics for
  model-based deep reinforcement learning with model-free fine-tuning, 2018
  IEEE International Conference on Robotics and Automation (ICRA) (2018)
  7559--7566.

\bibitem{watkins1989learning}
C.~J. C.~H. Watkins, Learning from delayed rewards, Ph.D. thesis, King's
  College (1989).

\bibitem{sutton2018reinforcement}
R.~S. Sutton, A.~G. Barto, Reinforcement learning: An introduction, MIT press,
  2018.

\bibitem{tsitsiklis1994asynchronous}
J.~N. Tsitsiklis, Asynchronous stochastic approximation and q-learning, Machine
  learning 16~(3) (1994) 185--202.

\bibitem{qu2020finite}
G.~Qu, A.~Wierman, Finite-time analysis of asynchronous stochastic
  approximation and $ q $-learning, Conference on Learning Theory (2020)
  3185--3205.

\bibitem{watkins1992q}
C.~J. Watkins, P.~Dayan, Q-learning, Machine learning 8~(3) (1992) 279--292.

\bibitem{rummery1994line}
G.~A. Rummery, M.~Niranjan, On-line Q-learning using connectionist systems,
  Vol.~37, Citeseer, 1994.

\bibitem{mnih2015human}
V.~Mnih, K.~Kavukcuoglu, D.~Silver, A.~A. Rusu, J.~Veness, M.~G. Bellemare,
  A.~Graves, M.~Riedmiller, A.~K. Fidjeland, G.~Ostrovski, et~al., Human-level
  control through deep reinforcement learning, nature 518~(7540) (2015)
  529--533.

\bibitem{lin1992self}
L.-J. Lin, Self-improving reactive agents based on reinforcement learning,
  planning and teaching, Machine learning 8~(3) (1992) 293--321.

\bibitem{gu2016continuous}
S.~Gu, T.~Lillicrap, I.~Sutskever, S.~Levine, Continuous deep q-learning with
  model-based acceleration, International conference on machine learning (2016)
  2829--2838.

\bibitem{van2016deep}
H.~Van~Hasselt, A.~Guez, D.~Silver, Deep reinforcement learning with double
  q-learning, Proceedings of the AAAI conference on artificial intelligence
  30~(1) (2016).

\bibitem{wang2016dueling}
Z.~Wang, T.~Schaul, M.~Hessel, H.~Hasselt, M.~Lanctot, N.~Freitas, Dueling
  network architectures for deep reinforcement learning, International
  conference on machine learning (2016) 1995--2003.

\bibitem{duan2021distributional}
J.~Duan, Y.~Guan, S.~E. Li, Y.~Ren, Q.~Sun, B.~Cheng, Distributional soft
  actor-critic: Off-policy reinforcement learning for addressing value
  estimation errors, IEEE Transactions on Neural Networks and Learning Systems
  (2021).

\bibitem{sutton1999policy}
R.~S. Sutton, D.~McAllester, S.~Singh, Y.~Mansour, Policy gradient methods for
  reinforcement learning with function approximation, Advances in neural
  information processing systems 12 (1999).

\bibitem{williams1992simple}
R.~J. Williams, Simple statistical gradient-following algorithms for
  connectionist reinforcement learning, Machine learning 8~(3) (1992) 229--256.

\bibitem{silver2014deterministic}
D.~Silver, G.~Lever, N.~Heess, T.~Degris, D.~Wierstra, M.~Riedmiller,
  Deterministic policy gradient algorithms, International conference on machine
  learning (2014) 387--395.

\bibitem{lillicrap2015continuous}
T.~P. Lillicrap, J.~J. Hunt, A.~Pritzel, N.~Heess, T.~Erez, Y.~Tassa,
  D.~Silver, D.~Wierstra, Continuous control with deep reinforcement learning,
  ICRA (Poster) (2016).

\bibitem{schulman2015trust}
J.~Schulman, S.~Levine, P.~Abbeel, M.~Jordan, P.~Moritz, Trust region policy
  optimization, International conference on machine learning (2015) 1889--1897.

\bibitem{schulman2017proximal}
J.~Schulman, F.~Wolski, P.~Dhariwal, A.~Radford, O.~Klimov, Proximal policy
  optimization algorithms, arXiv preprint arXiv:1707.06347 (2017).

\bibitem{ke2017lightgbm}
G.~Ke, Q.~Meng, T.~Finley, T.~Wang, W.~Chen, W.~Ma, Q.~Ye, T.-Y. Liu, Lightgbm:
  A highly efficient gradient boosting decision tree, Advances in neural
  information processing systems 30 (2017) 3146--3154.

\bibitem{mnih2016asynchronous}
V.~Mnih, A.~P. Badia, M.~Mirza, A.~Graves, T.~Lillicrap, T.~Harley, D.~Silver,
  K.~Kavukcuoglu, Asynchronous methods for deep reinforcement learning,
  International conference on machine learning (2016) 1928--1937.

\bibitem{wang2016learning}
J.~X. Wang, Z.~Kurth-Nelson, D.~Tirumala, H.~Soyer, J.~Z. Leibo, R.~Munos,
  C.~Blundell, D.~Kumaran, M.~Botvinick, Learning to reinforcement learn, arXiv
  preprint arXiv:1611.05763 (2016).

\bibitem{haarnoja2018soft}
T.~Haarnoja, A.~Zhou, P.~Abbeel, S.~Levine, Soft actor-critic: Off-policy
  maximum entropy deep reinforcement learning with a stochastic actor,
  International conference on machine learning (2018) 1861--1870.

\bibitem{gronauer2022multi}
S.~Gronauer, K.~Diepold, Multi-agent deep reinforcement learning: a survey,
  Artificial Intelligence Review 55~(2) (2022) 895--943.

\bibitem{zhang2021multi}
K.~Zhang, Z.~Yang, T.~Ba{\c{s}}ar, Multi-agent reinforcement learning: A
  selective overview of theories and algorithms, Handbook of Reinforcement
  Learning and Control (2021) 321--384.

\bibitem{omidshafiei2017deep}
S.~Omidshafiei, J.~Pazis, C.~Amato, J.~P. How, J.~Vian, Deep decentralized
  multi-task multi-agent reinforcement learning under partial observability,
  International Conference on Machine Learning (2017) 2681--2690.

\bibitem{wang20213m}
W.~Wang, Y.~Liu, R.~Srikant, L.~Ying, 3m-rl: Multi-resolution, multi-agent,
  mean-field reinforcement learning for autonomous uav routing, IEEE
  Transactions on Intelligent Transportation Systems (2021).

\bibitem{tan1993multi}
M.~Tan, Multi-agent reinforcement learning: Independent vs. cooperative agents,
  Proceedings of the tenth international conference on machine learning (1993)
  330--337.

\bibitem{li2023ace}
C.~Li, J.~Liu, Y.~Zhang, Y.~Wei, Y.~Niu, Y.~Yang, Y.~Liu, W.~Ouyang, Ace:
  Cooperative multi-agent q-learning with bidirectional action-dependency,
  Proceedings of the AAAI conference on artificial intelligence 37~(7) (2023)
  8536--8544.

\bibitem{matignon2007hysteretic}
L.~Matignon, G.~J. Laurent, N.~Le~Fort-Piat, Hysteretic q-learning: an
  algorithm for decentralized reinforcement learning in cooperative multi-agent
  teams, 2007 IEEE/RSJ International Conference on Intelligent Robots and
  Systems (2007) 64--69.

\bibitem{foerster2018counterfactual}
J.~Foerster, G.~Farquhar, T.~Afouras, N.~Nardelli, S.~Whiteson, Counterfactual
  multi-agent policy gradients, Proceedings of the AAAI conference on
  artificial intelligence 32~(1) (2018).

\bibitem{li2019robust}
S.~Li, Y.~Wu, X.~Cui, H.~Dong, F.~Fang, S.~Russell, Robust multi-agent
  reinforcement learning via minimax deep deterministic policy gradient,
  Proceedings of the AAAI conference on artificial intelligence 33~(1) (2019)
  4213--4220.

\bibitem{yang2018mean}
Y.~Yang, R.~Luo, M.~Li, M.~Zhou, W.~Zhang, J.~Wang, Mean field multi-agent
  reinforcement learning, International conference on machine learning (2018)
  5571--5580.

\bibitem{sunehag2017value}
P.~Sunehag, G.~Lever, A.~Gruslys, W.~M. Czarnecki, V.~Zambaldi, M.~Jaderberg,
  M.~Lanctot, N.~Sonnerat, J.~Z. Leibo, K.~Tuyls, et~al., Value-decomposition
  networks for cooperative multi-agent learning, arXiv preprint
  arXiv:1706.05296 (2017).

\bibitem{rashid2020weighted}
T.~Rashid, G.~Farquhar, B.~Peng, S.~Whiteson, Weighted qmix: Expanding
  monotonic value function factorisation for deep multi-agent reinforcement
  learning, Advances in neural information processing systems 33 (2020)
  10199--10210.

\bibitem{schmidt2017review}
M.~Schmidt, A review of aircraft turnaround operations and simulations,
  Progress in Aerospace Sciences 92 (2017) 25--38.

\bibitem{ahmed2018cooperative}
M.~S. Ahmed, S.~Alam, M.~Barlow, A cooperative co-evolutionary optimisation
  model for best-fit aircraft sequence and feasible runway configuration in a
  multi-runway airport, Aerospace 5~(3) (2018) 85.

\bibitem{conde2016trajectory}
M.~Conde Rocha~Murca, R.~DeLaura, R.~J. Hansman, R.~Jordan, T.~Reynolds,
  H.~Balakrishnan, Trajectory clustering and classification for
  characterization of air traffic flows, 16th AIAA Aviation Technology,
  Integration, and Operations Conference (2016) 3760.

\bibitem{lee2016taxi}
H.~Lee, W.~Malik, Y.~C. Jung, Taxi-out time prediction for departures at
  charlotte airport using machine learning techniques, 16th AIAA Aviation
  Technology, Integration, and Operations Conference (2016) 3910.

\bibitem{takeichi2017prediction}
N.~Takeichi, R.~Kaida, A.~Shimomura, T.~Yamauchi, Prediction of delay due to
  air traffic control by machine learning, AIAA modeling and simulation
  technologies conference (2017) 1323.

\bibitem{choi2016prediction}
S.~Choi, Y.~J. Kim, S.~Briceno, D.~Mavris, Prediction of weather-induced
  airline delays based on machine learning algorithms, 2016 IEEE/AIAA 35th
  Digital Avionics Systems Conference (DASC) (2016) 1--6.

\bibitem{ayhan2016aircraft}
S.~Ayhan, H.~Samet, Aircraft trajectory prediction made easy with predictive
  analytics, Proceedings of the 22nd ACM SIGKDD International Conference on
  Knowledge Discovery and Data Mining (2016) 21--30.

\bibitem{alligier2015machine}
R.~Alligier, D.~Gianazza, N.~Durand, Machine learning and mass estimation
  methods for ground-based aircraft climb prediction, IEEE Transactions on
  Intelligent Transportation Systems 16~(6) (2015) 3138--3149.

\bibitem{OTDelay}
U.~D.~O. TRANSPORTATION,
  \href{https://www.transtats.bts.gov/OT_Delay/OT_DelayCause1.asp}{Bureau of
  transportation statistics}.
\newline\urlprefix\url{https://www.transtats.bts.gov/OT_Delay/OT_DelayCause1.asp}

\bibitem{ACASX}
d.~EKim, S.~Bak,
  \href{https://github.com/stanleybak/acasxu_closed_loop_sim}{Acasxu closed
  loop simulation falsification benchmark}.
\newline\urlprefix\url{https://github.com/stanleybak/acasxu_closed_loop_sim}

\bibitem{eurocontrol}
EUROCONTROL, \href{https://www.eurocontrol.int/ddr}{Ademand data repository}.
\newline\urlprefix\url{https://www.eurocontrol.int/ddr}

\bibitem{hoekstra2016bluesky}
J.~M. Hoekstra, J.~Ellerbroek, Bluesky atc simulator project: an open data and
  open source approach, Proceedings of the 7th international conference on
  research in air transportation 131 (2016) 132.

\bibitem{berndt2004jsbsim}
J.~Berndt, Jsbsim: An open source flight dynamics model in c++, AIAA Modeling
  and Simulation Technologies Conference and Exhibit (2004) 4923.

\bibitem{flightgear}
\href{https://www.flightgear.org/}{Flightgear flight simulator sophisticated,
  professional, open-source}.
\newline\urlprefix\url{https://www.flightgear.org/}

\bibitem{wang2022review}
Z.~Wang, W.~Pan, H.~Li, X.~Wang, Q.~Zuo, Review of deep reinforcement learning
  approaches for conflict resolution in air traffic control, Aerospace 9~(6)
  (2022) 294.

\bibitem{harman1989tcas}
W.~H. Harman, Tcas- a system for preventing midair collisions, The Lincoln
  Laboratory Journal 2~(3) (1989) 437--457.

\bibitem{jeannin2015formal}
J.-B. Jeannin, K.~Ghorbal, Y.~Kouskoulas, R.~Gardner, A.~Schmidt, E.~Zawadzki,
  A.~Platzer, Formal verification of acas x, an industrial airborne collision
  avoidance system, 2015 International Conference on Embedded Software (EMSOFT)
  (2015) 127--136.

\bibitem{kochenderfer2012next}
M.~J. Kochenderfer, J.~E. Holland, J.~P. Chryssanthacopoulos, Next-generation
  airborne collision avoidance system, Tech. rep., Massachusetts Institute of
  Technology-Lincoln Laboratory Lexington United States (2012).

\bibitem{wulfe2017uav}
B.~Wulfe, Uav collision avoidance policy optimization with deep reinforcement
  learning (2017).

\bibitem{hermans2021towards}
M.~Herman, Towards explainable automation for air traffic control using deep
  q-learning from demonstrations and reward decomposition, Master's thesis,
  Aerospace Engineering (2021).

\bibitem{brittain2019autonomous}
M.~Brittain, P.~Wei, Autonomous separation assurance in an high-density en
  route sector: A deep multi-agent reinforcement learning approach, 2019 IEEE
  Intelligent Transportation Systems Conference (ITSC) (2019) 3256--3262.

\bibitem{brittain2020deep}
M.~Brittain, X.~Yang, P.~Wei, A deep multi-agent reinforcement learning
  approach to autonomous separation assurance, AIAA Journal of Aerospace
  Information Systems 18~(12) (2021).

\bibitem{brittain2021one}
M.~W. Brittain, P.~Wei, One to any: Distributed conflict resolution with deep
  multi-agent reinforcement learning and long short-term memory, AIAA Scitech
  2021 Forum (2021) 1952.

\bibitem{guo2021safety}
W.~Guo, M.~Brittain, P.~Wei, Safety enhancement for deep reinforcement learning
  in autonomous separation assurance, 2021 IEEE International Intelligent
  Transportation Systems Conference (ITSC) (2021) 348--354.

\bibitem{hu2021obstacle}
J.~Hu, X.~Yang, W.~Wang, P.~Wei, L.~Ying, Y.~Liu, Obstacle avoidance for uas in
  continuous action space using deep reinforcement learning, IEEE Access 10
  (2022) 90623--90634.

\bibitem{pham2019reinforcement}
D.-T. Pham, N.~P. Tran, S.~K. Goh, S.~Alam, V.~Duong, Reinforcement learning
  for two-aircraft conflict resolution in the presence of uncertainty, 2019
  IEEE-RIVF International Conference on Computing and Communication
  Technologies (RIVF) (2019) 1--6.

\bibitem{wang2019deep}
Z.~Wang, H.~Li, J.~Wang, F.~Shen, Deep reinforcement learning based conflict
  detection and resolution in air traffic control, IET Intelligent Transport
  Systems 13~(6) (2019) 1041--1047.

\bibitem{li2019optimizing}
S.~Li, M.~Egorov, M.~Kochenderfer, Optimizing collision avoidance in dense
  airspace using deep reinforcement learning, Thirteenth USA/Europe Air Traffic
  Management Research and Development Seminar (ATM) (2019).

\bibitem{bertram2020distributed}
J.~Bertram, P.~Wei, Distributed computational guidance for high-density urban
  air mobility with cooperative and non-cooperative collision avoidance, AIAA
  Scitech 2020 Forum (2020) 1371.

\bibitem{brittain2018towards}
M.~W. Brittain, P.~Wei, Towards autonomous air trac control for sequencing and
  separation-a deep reinforcement learning approach, 2018 Aviation Technology,
  Integration, and Operations Conference (2018) 3664.

\bibitem{jacob2022autonomous}
B.~Jacob, A.~Kaushik, P.~Velavan, M.~Sharma, Autonomous drones for medical
  assistance using reinforcement learning, Advances in Augmented Reality and
  Virtual Reality (2022) 133--156.

\bibitem{xu2021method}
Q.~Xu, J.~Huang, Z.~Liu, H.~Ding, A method based on deep reinforcement learning
  to generate control strategy for aircrafts in terminal sector, Artificial
  Intelligence in China (2021) 356--363.

\bibitem{garcia2023isuam}
C.~P. Garcia, L.~Weigang, N.~S. Hirata, C.~Neumann, Isuam: Intelligent and safe
  uam with deep reinforcement learning, 2023 IEEE 29th International Conference
  on Parallel and Distributed Systems (ICPADS) (2023) 378--383.

\bibitem{gal2016dropout}
Y.~Gal, Z.~Ghahramani, Dropout as a bayesian approximation: Representing model
  uncertainty in deep learning, international conference on machine learning
  (2016) 1050--1059.

\bibitem{hu2020uas}
J.~Hu, Y.~Liu, Uas conflict resolution integrating a risk-based operational
  safety bound as airspace reservation with reinforcement learning, AIAA
  Scitech 2020 Forum (2020) 1372.

\bibitem{dalmauair}
R.~Dalmau, E.~Allard, Air traffic control using message passing neural networks
  and multi-agent reinforcement learning, Proceedings of the 10th SESAR
  Innovation Days, Virtual Event (2020) 7--10.

\bibitem{zhao2021physics}
P.~Zhao, Y.~Liu, Physics informed deep reinforcement learning for aircraft
  conflict resolution, IEEE Transactions on Intelligent Transportation Systems
  (2021).

\bibitem{panoutsakopoulos2022towards}
C.~Panoutsakopoulos, B.~Yuksek, G.~Inalhan, A.~Tsourdos, Towards safe deep
  reinforcement learning for autonomous airborne collision avoidance systems,
  AIAA SCITECH 2022 Forum (2022) 2102.

\bibitem{tran2019intelligent}
N.~P. Tran, D.-T. Pham, S.~K. Goh, S.~Alam, V.~Duong, An intelligent
  interactive conflict solver incorporating air traffic controllers'
  preferences using reinforcement learning, 2019 Integrated Communications,
  Navigation and Surveillance Conference (ICNS) (2019) 1--8.

\bibitem{tran2020interactive}
P.~N. Tran, D.-T. Pham, S.~K. Goh, S.~Alam, V.~Duong, An interactive conflict
  solver for learning air traffic conflict resolutions, Journal of Aerospace
  Information Systems 17~(6) (2020) 271--277.

\bibitem{wen2019application}
H.~Wen, H.~Li, Z.~Wang, X.~Hou, K.~He, Application of ddpg-based collision
  avoidance algorithm in air traffic control, 2019 12th International Symposium
  on Computational Intelligence and Design (ISCID) 1 (2019) 130--133.

\bibitem{pham2019machine}
D.-T. Pham, N.~P. Tran, S.~Alam, V.~Duong, D.~Delahaye, A machine learning
  approach for conflict resolution in dense traffic scenarios with
  uncertainties, ATM Seminar 2019, 13th USA/Europe ATM R\&D Seminar (2019).

\bibitem{ribeiro2020determining}
M.~Ribeiro, J.~Ellerbroek, J.~Hoekstra, Determining optimal conflict avoidance
  manoeuvres at high densities with reinforcement learning, Proceedings of the
  Tenth SESAR Innovation Days, Virtual Conference (2020) 7--10.

\bibitem{isufaj2021towards}
R.~Isufaj, D.~Aranega~Sebastia, M.~A. Piera, Towards conflict resolution with
  deep multi-agent reinforcement learning, Proceedings of the 14th USA/Europe
  Air Traffic Management Research and Development Seminar (ATM2021), New
  Orleans, LA, USA (2021) 20--24.

\bibitem{lai2021multi}
J.~Lai, K.~Cai, Z.~Liu, Y.~Yang, A multi-agent reinforcement learning approach
  for conflict resolution in dense traffic scenarios, 2021 IEEE/AIAA 40th
  Digital Avionics Systems Conference (DASC) (2021) 1--9.

\bibitem{mollinga2020autonomous}
J.~Mollinga, H.~van Hoof, An autonomous free airspace en-route controller using
  deep reinforcement learning techniques, International Conference for Research
  in Air Transportation (ICRAT) (2020).

\bibitem{zu2021multi}
W.~Zu, H.~Yang, R.~Liu, Y.~Ji, A multi-dimensional goal aircraft guidance
  approach based on reinforcement learning with a reward shaping algorithm,
  Sensors 21~(16) (2021) 5643.

\bibitem{zhao2021reinforcement}
Y.~Zhao, J.~Guo, C.~Bai, H.~Zheng, Reinforcement learning-based collision
  avoidance guidance algorithm for fixed-wing uavs, Complexity 2021 (2021).

\bibitem{alvarez2023towards}
L.~E. Alvarez, M.~Brittain, K.~Breeden, Towards a standardized reinforcement
  learning framework for aam contingency management, 2023 IEEE/AIAA 42nd
  Digital Avionics Systems Conference (DASC) (2023) 1--6.

\bibitem{brittain2024improving}
M.~W. Brittain, L.~E. Alvarez, K.~Breeden, Improving autonomous separation
  assurance through distributed reinforcement learning with attention networks,
  Proceedings of the AAAI Conference on Artificial Intelligence 38~(21) (2024)
  22857--22863.

\bibitem{yang2020scalable}
X.~Yang, P.~Wei, Scalable multi-agent computational guidance with separation
  assurance for autonomous urban air mobility, Journal of Guidance, Control,
  and Dynamics 43~(8) (2020) 1473--1486.

\bibitem{chaslot2008monte}
G.~Chaslot, S.~Bakkes, I.~Szita, P.~Spronck, Monte-carlo tree search: A new
  framework for game ai, Proceedings of the AAAI Conference on Artificial
  Intelligence and Interactive Digital Entertainment 4~(1) (2021) 216--217.

\bibitem{schrittwieser2020mastering}
J.~Schrittwieser, I.~Antonoglou, T.~Hubert, K.~Simonyan, L.~Sifre, S.~Schmitt,
  A.~Guez, E.~Lockhart, D.~Hassabis, T.~Graepel, et~al., Mastering atari, go,
  chess and shogi by planning with a learned model, Nature 588~(7839) (2020)
  604--609.

\bibitem{yilmaz2021deep}
E.~Yilmaz, O.~Sanni, M.~Kotwicz~Herniczek, B.~German, Deep reinforcement
  learning approach to air traffic optimization using the muzero algorithm,
  AIAA AVIATION 2021 FORUM (2021) 2377.

\bibitem{singh2021approximate}
A.~J. Singh, A.~Kumar, H.~C. Lau, Approximate difference rewards for scalable
  multigent reinforcement learning, Proceedings of the 20th International
  Conference on Autonomous Agents and MultiAgent Systems (2021) 1655--1657.

\bibitem{isufaj2022multi}
R.~Isufaj, M.~Omeri, M.~A. Piera, Multi-uav conflict resolution with graph
  convolutional reinforcement learning, Applied Sciences 12~(2) (2022) 610.

\bibitem{deniz2024reinforcement}
S.~Deniz, Y.~Wu, Y.~Shi, Z.~Wang, A reinforcement learning approach to vehicle
  coordination for structured advanced air mobility, Green Energy and
  Intelligent Transportation (2024) 100157.

\bibitem{julian2016policy}
K.~D. Julian, J.~Lopez, J.~S. Brush, M.~P. Owen, M.~J. Kochenderfer, Policy
  compression for aircraft collision avoidance systems, 2016 IEEE/AIAA 35th
  Digital Avionics Systems Conference (DASC) (2016) 1--10.

\bibitem{CRUCIOL2013141}
L.~L. Cruciol, A.~C. de~Arruda~Jr, L.~Weigang, L.~Li, A.~M. Crespo, Reward
  functions for learning to control in air traffic flow management,
  Transportation Research Part C: Emerging Technologies 35 (2013) 141--155.

\bibitem{xu2020synchronised}
Y.~Xu, X.~Prats, D.~Delahaye, Synchronised demand-capacity balancing in
  collaborative air traffic flow management, Transportation Research Part C:
  Emerging Technologies 114 (2020) 359--376.

\bibitem{9594397}
C.~Huang, Y.~Xu, Integrated frameworks of unsupervised, supervised and
  reinforcement learning for solving air traffic flow management problem, 2021
  IEEE/AIAA 40th Digital Avionics Systems Conference (DASC) (2021) 1--10\href
  {https://doi.org/10.1109/DASC52595.2021.9594397}
  {\path{doi:10.1109/DASC52595.2021.9594397}}.

\bibitem{xie2021reinforcement}
Y.~Xie, A.~Gardi, R.~Sabatini, Reinforcement learning-based flow management
  techniques for urban air mobility and dense low-altitude air traffic
  operations, 2021 IEEE/AIAA 40th Digital Avionics Systems Conference (DASC)
  (2021) 1--10.

\bibitem{tang2021multi}
Y.~Tang, Y.~Xu, Multi-agent deep reinforcement learning for solving large-scale
  air traffic flow management problem: A time-step sequential decision
  approach, 2021 IEEE/AIAA 40th Digital Avionics Systems Conference (DASC)
  (2021) 1--10.

\bibitem{kravaris2019resolving}
T.~Kravaris, C.~Spatharis, A.~Bastas, G.~A. Vouros, K.~Blekas, G.~Andrienko,
  N.~Andrienko, J.~M.~C. Garcia, Resolving congestions in the air traffic
  management domain via multiagent reinforcement learning methods, arXiv
  preprint arXiv:1912.06860 (2019).

\bibitem{spatharis2021hierarchical}
C.~Spatharis, A.~Bastas, T.~Kravaris, K.~Blekas, G.~A. Vouros, J.~M. Cordero,
  Hierarchical multiagent reinforcement learning schemes for air traffic
  management, Neural Computing and Applications (2021) 1--13.

\bibitem{duong2019decentralizing}
T.~Duong, K.~K. Todi, U.~Chaudhary, H.-L. Truong, Decentralizing air traffic
  flow management with blockchain-based reinforcement learning, 2019 IEEE 17th
  International Conference on Industrial Informatics (INDIN) 1 (2019)
  1795--1800.

\bibitem{chen2021demand}
Y.~Chen, Y.~Xu, M.~Hu, L.~Yang, Demand and capacity balancing technology based
  on multi-agent reinforcement learning, 2021 IEEE/AIAA 40th Digital Avionics
  Systems Conference (DASC) (2021) 1--9.

\bibitem{chen2024integrated}
S.~Chen, A.~D. Evans, M.~Brittain, P.~Wei, Integrated conflict management for
  uam with strategic demand capacity balancing and learning-based tactical
  deconfliction, IEEE Transactions on Intelligent Transportation Systems
  (2024).

\bibitem{groot2024analysis}
D.~Groot, J.~Ellerbroek, J.~Hoekstra, Analysis of the impact of traffic density
  on training of reinforcement learning based conflict resolution methods for
  drones, Engineering Applications of Artificial Intelligence 133 (2024)
  108066.

\bibitem{spatharis2018multiagent}
C.~Spatharis, T.~Kravaris, G.~A. Vouros, K.~Blekas, G.~Chalkiadakis, J.~M.~C.
  Garcia, E.~C. Fernandez, Multiagent reinforcement learning methods to resolve
  demand capacity balance problems, Proceedings of the 10th Hellenic Conference
  on Artificial Intelligence (2018) 1--9.

\bibitem{estes2017data}
A.~Estes, M.~Ball, Data-driven planning for ground delay programs,
  Transportation Research Record 2603~(1) (2017) 13--20.

\bibitem{james2021tmi}
J.~Jones, Z.~Ellenbogen, Y.~Glina, Recommending strategic air traffic
  management initiatives in convective weather, Fourteenth USA/Europe Air
  Traffic Management Research and Development Seminar (ATM2021), Lexington, MA
  02421, USA (2021).

\bibitem{bloem2015ground}
M.~Bloem, N.~Bambos, Ground delay program analytics with behavioral cloning and
  inverse reinforcement learning, Journal of Aerospace Information Systems
  12~(3) (2015) 299--313.

\bibitem{george2015reinforcement}
E.~George, S.~S. Khan, Reinforcement learning for taxi-out time prediction: An
  improved q-learning approach, 2015 International Conference on Computing and
  Network Communications (CoCoNet) (2015) 757--764.

\bibitem{memarzadeh2023airport}
M.~Memarzadeh, T.~G. Puranik, K.~M. Kalyanam, W.~Ryan, Airport runway
  configuration management with offline model-free reinforcement learning, AIAA
  SciTech 2023 Forum (2023) 0504.

\bibitem{nethi2024optimization}
S.~Nethi, M.~Memarzadeh, K.~Kalyanam, Optimization of runway configurations
  with forecast-augmented offline reinforcement learning, AIAA SCITECH 2024
  Forum (2024) 0533.

\bibitem{talluri2004theory}
K.~T. Talluri, G.~Van~Ryzin, G.~Van~Ryzin, The theory and practice of revenue
  management, Vol.~1, Springer, 2004.

\bibitem{gosavii2002reinforcement}
A.~Gosavii, N.~Bandla, T.~K. Das, A reinforcement learning approach to a single
  leg airline revenue management problem with multiple fare classes and
  overbooking, IIE transactions 34~(9) (2002) 729--742.

\bibitem{lawhead2019bounded}
R.~J. Lawhead, A.~Gosavi, A bounded actor--critic reinforcement learning
  algorithm applied to airline revenue management, Engineering Applications of
  Artificial Intelligence 82 (2019) 252--262.

\bibitem{bondoux2020reinforcement}
N.~Bondoux, A.~Q. Nguyen, T.~Fiig, R.~Acuna-Agost, Reinforcement learning
  applied to airline revenue management, Journal of Revenue and Pricing
  Management 19~(5) (2020) 332--348.

\bibitem{shihab2021deep}
S.~A. Shihab, P.~Wei, A deep reinforcement learning approach to seat inventory
  control for airline revenue management, Journal of Revenue and Pricing
  Management (2021) 1--17.

\bibitem{wang2021solving}
R.~Wang, X.~Gan, Q.~Li, X.~Yan, Solving a joint pricing and inventory control
  problem for perishables via deep reinforcement learning, Complexity 2021
  (2021).

\bibitem{alamdari2021deep}
N.~E. Alamdari, G.~Savard, Deep reinforcement learning in seat inventory
  control problem: an action generation approach, Journal of Revenue and
  Pricing Management 20~(5) (2021) 566--579.

\bibitem{belobaba1987air}
P.~Belobaba, Air travel demand and airline seat inventory management, Ph.D.
  thesis, Massachusetts Institute of Technology (1987).

\bibitem{Emami_2022}
S.~A. Emami, P.~Castaldi, A.~Banazadeh,
  \href{http://dx.doi.org/10.1016/j.arcontrol.2022.04.006}{Neural network-based
  flight control systems: Present and future}, Annual Reviews in Control 53
  (2022) 97–137.
\newblock \href {https://doi.org/10.1016/j.arcontrol.2022.04.006}
  {\path{doi:10.1016/j.arcontrol.2022.04.006}}.
\newline\urlprefix\url{http://dx.doi.org/10.1016/j.arcontrol.2022.04.006}

\bibitem{li2022adaptive}
Z.~Li, X.~Chen, M.~Xie, Z.~Zhao, Adaptive fault-tolerant tracking control of
  flying-wing unmanned aerial vehicle with system input saturation and state
  constraints, Transactions of the Institute of Measurement and Control 44~(4)
  (2022) 880--891.

\bibitem{li2021maneuvering}
B.~Li, Z.-p. Yang, D.-q. Chen, S.-y. Liang, H.~Ma, Maneuvering target tracking
  of uav based on mn-ddpg and transfer learning, Defence Technology 17~(2)
  (2021) 457--466.

\bibitem{zhao2020research}
W.~Zhao, H.~Chu, X.~Miao, L.~Guo, H.~Shen, C.~Zhu, F.~Zhang, D.~Liang, Research
  on the multiagent joint proximal policy optimization algorithm controlling
  cooperative fixed-wing uav obstacle avoidance, Sensors 20~(16) (2020) 4546.

\bibitem{lee2018vision}
S.~Lee, T.~Shim, S.~Kim, J.~Park, K.~Hong, H.~Bang, Vision-based autonomous
  landing of a multi-copter unmanned aerial vehicle using reinforcement
  learning, 2018 International Conference on Unmanned Aircraft Systems (ICUAS)
  (2018) 108--114.

\bibitem{xian2021robust}
B.~Xian, X.~Zhang, H.~Zhang, X.~Gu, Robust adaptive control for a small
  unmanned helicopter using reinforcement learning, IEEE Transactions on Neural
  Networks and Learning Systems (2021).

\bibitem{zhen2020deep}
Y.~Zhen, M.~Hao, W.~Sun, Deep reinforcement learning attitude control of
  fixed-wing uavs, 2020 3rd International Conference on Unmanned Systems (ICUS)
  (2020) 239--244.

\bibitem{huang2019attitude}
X.~Huang, W.~Luo, J.~Liu, Attitude control of fixed-wing uav based on ddqn,
  2019 Chinese Automation Congress (CAC) (2019) 4722--4726.

\bibitem{huang2020model}
D.~Huang, J.~Hu, Z.~Peng, B.~Chen, M.~Hao, B.~K. Ghosh, Model-free based
  reinforcement learning control strategy of aircraft attitude systems, 2020
  Chinese Automation Congress (CAC) (2020) 743--748.

\bibitem{shi2016decoupled}
H.~Shi, X.~Li, K.-S. Hwang, W.~Pan, G.~Xu, Decoupled visual servoing with fuzzy
  q-learning, IEEE Transactions on Industrial Informatics 14~(1) (2016)
  241--252.

\bibitem{kim2017reinforcement}
D.~Kim, G.~Oh, Y.~Seo, Y.~Kim, Reinforcement learning-based optimal flat spin
  recovery for unmanned aerial vehicle, Journal of Guidance, Control, and
  Dynamics 40~(4) (2017) 1076--1084.

\bibitem{zuo2019flight}
Y.~Zuo, K.~Deng, Y.~Yang, T.~Huang, Flight attitude simulator control system
  design based on model-free reinforcement learning method, 2019 IEEE 3rd
  Advanced Information Management, Communicates, Electronic and Automation
  Control Conference (IMCEC) (2019) 355--361.

\bibitem{wang2020attitude}
Z.~Wang, W.~Luo, Q.~Gong, Y.~Cui, R.~Tao, Q.~Wang, Q.~Liang, S.~Wang, Attitude
  controller design based on deep reinforcement learning for low-cost aircraft,
  2020 Chinese Automation Congress (CAC) (2020) 463--467.

\bibitem{zhang2020autonomous}
K.~Zhang, K.~Li, H.~Shi, Z.~Zhang, Z.~Liu, Autonomous guidance maneuver control
  and decision-making algorithm based on deep reinforcement learning uav route,
  Systems Engineering and Electronics 42~(7) (2020) 1567--1574.

\bibitem{al2019machine}
M.~Al-Gabalawy, Machine learning for aircraft control, Journal of Advanced
  Research in Dynamical and Control Systems 11 (2019) 3165--3191.

\bibitem{bohn2019deep}
E.~B{\o}hn, E.~M. Coates, S.~Moe, T.~A. Johansen, Deep reinforcement learning
  attitude control of fixed-wing uavs using proximal policy optimization, 2019
  International Conference on Unmanned Aircraft Systems (ICUAS) (2019)
  523--533.

\bibitem{wada2021unmanned}
D.~Wada, S.~A. Araujo-Estrada, S.~Windsor, Unmanned aerial vehicle pitch
  control under delay using deep reinforcement learning with continuous action
  in wind tunnel test, Aerospace 8~(9) (2021) 258.

\bibitem{blanke2006diagnosis}
M.~Blanke, M.~Kinnaert, J.~Lunze, M.~Staroswiecki, J.~Schr{\"o}der, Diagnosis
  and fault-tolerant control, Vol.~2, Springer, 2006.

\bibitem{koch2019reinforcement}
W.~Koch, R.~Mancuso, R.~West, A.~Bestavros, Reinforcement learning for uav
  attitude control, ACM Transactions on Cyber-Physical Systems 3~(2) (2019)
  1--21.

\bibitem{wang2019deterministic}
Y.~Wang, J.~Sun, H.~He, C.~Sun, Deterministic policy gradient with integral
  compensator for robust quadrotor control, IEEE Transactions on Systems, Man,
  and Cybernetics: Systems 50~(10) (2019) 3713--3725.

\bibitem{fei2020learn}
F.~Fei, Z.~Tu, D.~Xu, X.~Deng, Learn-to-recover: Retrofitting uavs with
  reinforcement learning-assisted flight control under cyber-physical attacks,
  2020 IEEE International Conference on Robotics and Automation (ICRA) (2020)
  7358--7364.

\bibitem{9595275}
L.~Bhan, M.~Quinones-Grueiro, G.~Biswas, Fault tolerant control combining
  reinforcement learning and model-based control, 2021 5th International
  Conference on Control and Fault-Tolerant Systems (SysTol) (2021) 31--36.

\bibitem{wickman2021exploring}
A.~Wickman, Exploring feasibility of reinforcement learning flight route
  planning, Bachelor's thesis, Department of Computer and Information Science
  (2021).

\bibitem{azar2021drone}
A.~T. Azar, A.~Koubaa, N.~Ali~Mohamed, H.~A. Ibrahim, Z.~F. Ibrahim, M.~Kazim,
  A.~Ammar, B.~Benjdira, A.~M. Khamis, I.~A. Hameed, et~al., Drone deep
  reinforcement learning: A review, Electronics 10~(9) (2021) 999.

\bibitem{ragi2013uav}
S.~Ragi, E.~K. Chong, Uav path planning in a dynamic environment via partially
  observable markov decision process, IEEE Transactions on Aerospace and
  Electronic Systems 49~(4) (2013) 2397--2412.

\bibitem{zhang2015geometric}
B.~Zhang, Z.~Mao, W.~Liu, J.~Liu, Geometric reinforcement learning for path
  planning of uavs, Journal of Intelligent \& Robotic Systems 77~(2) (2015)
  391--409.

\bibitem{yan2020towards}
C.~Yan, X.~Xiang, C.~Wang, Towards real-time path planning through deep
  reinforcement learning for a uav in dynamic environments, Journal of
  Intelligent \& Robotic Systems 98~(2) (2020) 297--309.

\bibitem{tumer2007distributed}
K.~Tumer, A.~Agogino, Distributed agent-based air traffic flow management,
  Proceedings of the 6th international joint conference on Autonomous agents
  and multiagent systems (2007) 1--8.

\bibitem{huang2021integrated}
C.~Huang, Y.~Xu, Integrated frameworks of unsupervised, supervised and
  reinforcement learning for solving air traffic flow management problem, 2021
  IEEE/AIAA 40th Digital Avionics Systems Conference (DASC) (2021) 1--10.

\bibitem{bertram2021scalable}
J.~Bertram, P.~Wei, J.~Zambreno, Scalable fastmdp for pre-departure airspace
  reservation and strategic de-conflict, AIAA Scitech 2021 Forum (2021) 0779.

\bibitem{juntama2022hyperheuristic}
P.~Juntama, D.~Delahaye, S.~Chaimatanan, S.~Alam, Hyperheuristic approach based
  on reinforcement learning for air traffic complexity mitigation, AIAA Journal
  of Aerospace Information Systems (2022) 1--16.

\bibitem{zhang2021learning}
K.~Zhang, Y.~Yang, C.~Xu, D.~Liu, H.~Song, Learning-to-dispatch: Reinforcement
  learning based flight planning under emergency, 2021 IEEE International
  Intelligent Transportation Systems Conference (ITSC) (2021) 1821--1826.

\bibitem{ruan2021reinforcement}
J.~Ruan, Z.~Wang, F.~T. Chan, S.~Patnaik, M.~K. Tiwari, A reinforcement
  learning-based algorithm for the aircraft maintenance routing problem, Expert
  Systems with Applications 169 (2021) 114399.

\bibitem{deng2022lookahead}
Q.~Deng, B.~F. Santos, Lookahead approximate dynamic programming for stochastic
  aircraft maintenance check scheduling optimization, European Journal of
  Operational Research 299~(3) (2022) 814--833.

\bibitem{hu2021reinforcement}
Y.~Hu, X.~Miao, J.~Zhang, J.~Liu, E.~Pan, Reinforcement learning-driven
  maintenance strategy: A novel solution for long-term aircraft maintenance
  decision optimization, Computers \& Industrial Engineering 153 (2021) 107056.

\bibitem{ribeiro2022playful}
J.~Ribeiro, P.~Andrade, M.~Carvalho, C.~Silva, B.~Ribeiro, L.~Roque, Playful
  probes for design interaction with machine learning: A tool for aircraft
  condition-based maintenance planning and visualisation, Mathematics 10~(9)
  (2022) 1604.

\bibitem{dangut2022application}
M.~D. Dangut, I.~K. Jennions, S.~King, Z.~Skaf, Application of deep
  reinforcement learning for extremely rare failure prediction in aircraft
  maintenance, Mechanical Systems and Signal Processing 171 (2022) 108873.

\bibitem{van2017challenges}
P.~Van~Wesel, A.~E. Goodloe, Challenges in the verification of reinforcement
  learning algorithms, Tech. rep., NASA (2017).

\bibitem{baheri2022verification}
A.~Baheri, H.~Ren, B.~Johnson, P.~Razzaghi, P.~Wei, A verification framework
  for certifying learning-based safety-critical aviation systems, AIAA AVIATION
  2022 Forum (2022).
\newblock \href {https://doi.org/10.2514/6.2022-3965}
  {\path{doi:10.2514/6.2022-3965}}.

\bibitem{garcia2015comprehensive}
J.~Garc{\i}a, F.~Fern{\'a}ndez, A comprehensive survey on safe reinforcement
  learning, Journal of Machine Learning Research 16~(1) (2015) 1437--1480.

\bibitem{kiran2021deep}
B.~R. Kiran, I.~Sobh, V.~Talpaert, P.~Mannion, A.~A. Al~Sallab, S.~Yogamani,
  P.~P{\'e}rez, Deep reinforcement learning for autonomous driving: A survey,
  IEEE Transactions on Intelligent Transportation Systems (2021).

\bibitem{baheri2020deep}
A.~Baheri, S.~Nageshrao, H.~E. Tseng, I.~Kolmanovsky, A.~Girard, D.~Filev, Deep
  reinforcement learning with enhanced safety for autonomous highway driving,
  2020 IEEE Intelligent Vehicles Symposium (IV) (2020) 1550--1555.

\bibitem{baheri2022safe}
A.~Baheri, Safe reinforcement learning with mixture density network, with
  application to autonomous driving, Results in Control and Optimization 6
  (2022) 100095.

\bibitem{groot2021improving}
J.~Groot, Improving safety of vertical manoeuvres in a layered airspace with
  deep reinforcement learning, Master's thesis, Aerospace Engineering (2021).

\bibitem{lazarus2020runtime}
C.~Lazarus, J.~G. Lopez, M.~J. Kochenderfer, Runtime safety assurance using
  reinforcement learning, 2020 AIAA/IEEE 39th Digital Avionics Systems
  Conference (DASC) (2020) 1--9.

\bibitem{wu2022comparisons}
P.~Wu, J.~Chen, Comparisons of rrt and mcts for safe assured path planning in
  urban air mobility, AIAA SCITECH 2022 Forum (2022) 1841.

\bibitem{wu2022safety}
P.~Wu, X.~Yang, P.~Wei, J.~Chen, Safety assured online guidance with airborne
  separation for urban air mobility operations in uncertain environments, IEEE
  Transactions on Intelligent Transportation Systems (2022).

\bibitem{hawley2019real}
M.~Hawley, R.~Bharadwaj, V.~Venkataraman, Real-time mitigation of loss of
  separation events using reinforcement learning, 2019 IEEE/AIAA 38th Digital
  Avionics Systems Conference (DASC) (2019) 1--6.

\end{thebibliography}

\end{document}